\documentclass[aps,pre,showpacs,floatfix,longbibliography]{revtex4-1}
\usepackage{amsmath,amssymb,color,graphicx,array,bm,url}

\begin{document}

\title{Flocking turbulence of microswimmers in confined domains}

\author{L. Puggioni}
\affiliation{Dipartimento di Fisica and INFN, Universit\`a degli Studi di Torino, via P. Giuria 1, 10125 Torino, Italy.}

\author{G. Boffetta}
\affiliation{Dipartimento di Fisica and INFN, Universit\`a degli Studi di Torino, via P. Giuria 1, 10125 Torino, Italy.}

\author{S. Musacchio}
\thanks{Corresponding author}
\email{stefano.musacchio@unito.it}
\affiliation{Dipartimento di Fisica and INFN, Universit\`a degli Studi di Torino, via P. Giuria 1, 10125 Torino, Italy.}

\begin{abstract}
We extensively study the Toner-Tu-Swift-Hohenberg model of motile active 
matter by means of direct numerical simulations in a two-dimensional 
confined domain.
By exploring the space of parameters of the model we investigate the emergence of 
a new state of active turbulence which occurs when the
aligning interactions and the self-propulsion of the swimmers are strong. 
This regime of {\it flocking turbulence} is
characterized by a population of few strong vortices,
each surrounded by an island of coherent flocking motion. 
The energy spectrum of flocking turbulence displays a power-law scaling with
an exponent which depends weakly on the model parameters.
By increasing the confinement we observe that flocking turbulence becomes 
unstable: after a long transient, characterized by power-law distributed 
transition times, the system switches to the ordered state of a single 
{\it giant vortex}.
\end{abstract}

\date{\today}

\maketitle

\section{Introduction}
\label{sec1}
The study of motile active matter as an example of nonequilibrium
statistical system has become increasingly important in the last years 
\cite{marchetti2013hydrodynamics,ramaswamy2010mechanics}.
A wide class of active systems like bacterial
suspensions \cite{dombrowski2004self,wensink2012meso}, microtubule-kinesin
mixtures \cite{sanchez2012spontaneous,doostmohammadi2018active}, cell
tissues \cite{saw2017topological}, vibrated disks and
rods \cite{deseigne2010collective,aranson2007swirling}, Janus
particles \cite{nishiguchi2015mesoscopic} and motile
colloids \cite{bricard2013emergence}, exhibits a very rich phenomenology of
collective motions. MMany of them can develop 
a complex spatio-temporal chaos, called active turbulence 
for its resemblance to inertial turbulence in fluids
\cite{alert2021active}. 
To describe active turbulence a variety of continuum models has been
proposed \cite{shaebani2020computational,bar2020self} that can be divided in
polar models, characterized by a vectorial order parameter, and nematic models,
with a rank-2 tensorial order parameter \cite{bowick2022symmetry}.
In general, this continuum approach allows the use of hydrodynamical tools 
to investigate active matter dynamics.

The different order parameter
brings to peculiar phenomena. In particular, polar
systems can exhibit a long-range ordered phase, with the constituent particles
aligning to each other and moving with constant speed, a phenomenon known
as flocking \cite{ramaswamy2010mechanics}. Since the original works by
Vicsek \cite{vicsek1995novel} and Toner and Tu
\cite{toner1995long,toner2005hydrodynamics},
the investigation of flocking dynamics has attracted the attention of 
physicists, both from an experimental 
\cite{ballerini2008interaction,wioland2013confinement,bricard2013emergence}
and a theoretical
\cite{toner1998flocks,chate2008collective,toner2012reanalysis,cavagna2015flocking}
point of view.
In recent years, the Toner-Tu (TT) model
 has been subject of various generalization,
to incompressible flows
\cite{chen2015critical,chen2016mapping,chen2018incompressible},  
to curved surfaces
\cite{shankar2017topological}, 
to the high-Reynolds number limit
\cite{rana2020coarsening,rana2022phase} 
and to quenched disorder
\cite{toner2018hydrodynamic,PhysRevE.106.044608}. 

In particular, a modified version of the incompressible Toner-Tu model has been 
proposed to describe the dynamics of dense pusher-like microswimmers
\cite{wensink2012meso,dunkel2013fluid,dunkel2013minimal}.
In this version, denoted as Toner-Tu-Swift-Hohenberg (TTSH) model,
the Landau force which promotes the flocking
is combined with the Swift-Hohenberg operator for pattern
formation \cite{swift1977hydrodynamic}.
In two dimensions, the TTSH model exhibits a very rich phenomenology
which encompass different regimes: 
stationary vortex lattice \cite{james2018turbulence,reinken2022optimal}, 
isotropic mesoscale turbulence 
\cite{bratanov2015new,james2018vortex,PhysRevFluids.5.024302,xia2020lattice},
and turbulent crystal-like regime
\cite{james2018turbulence,james2021emergence}.
     
Linear stability analysis \cite{10.1007/978-3-319-66764-5_14}
predicts that the regime of uniform flocking
(which is present in the original Toner-Tu model)
is destabilized by the Swift-Hohenberg operator,
and therefore it cannot observed in the TTSH model.  
Nonetheless, recent works
\cite{mukherjee2021anomalous,singh2022lagrangian,mukherjee2022intermittency}.
have shown that, if the aligning potential is sufficiently strong,
the TTSH model displays the emergence of an inhomogeneous regime
characterized by the presence of large-scale, isolated vortices,
surrounded by regions of small vortices and elongated vortical structures,
called {\it vorticity streaks}.
Moreover, when the flow is confined in bounded domains, this state can evolve
into a regime of global circular flocking,
in which the swimmers are organized in a single, giant vortex
\cite{puggioni2022giant}.

In this paper, we present a detailed investigation
of the inhomogeneous regime of large-scale vortices.
We show that these structure arise from local attempts
to organize the flow in configurations of circular flocking
induced by the aligning potential.
The interactions between the flocking vortices
give rise to a new dynamical regime that we call {\it flocking turbulence}.  
By means of an extensive exploration of the parameter space 
we highlight the importance of the interplay between the Landau force
and the nonlinear advection term to induce the transition
from the regime of isotropic mesoscale turbulence
towards flocking turbulence. 

The remaining of this paper is organized as follows. 
In section~\ref{sec2} we introduce the model and the numerical method.
In section~\ref{sec3} we explore the parameters of the model and 
we report the condition under which the transition to flocking 
turbulence is observed. In section~\ref{sec4} we study the effect of
confinement and its role in the transition to the giant vortex
regime. Finally, section~\ref{sec5} is devoted to conclusions.

\section{Toner-Tu-Swift-Hohenberg model}
\label{sec2}

The Toner-Tu-Swift-Hohenberg (TTSH) model describes
the effective dynamics of a dense suspension of elongated pusher-like microswimmers
as a polar active fluid, governed by an incompressible Navier-Stokes-like
equation for the coarse-grained collective velocity
field ${\bm u}$:
\begin{equation}
  \partial_t {\bm u} + \lambda {\bm u} \cdot {\bm \nabla} {\bm u}
  = - {\bm \nabla} p - (\alpha + \beta |{\bm u}|^2
  + \Gamma_2 \nabla^2 + \Gamma_4 \nabla^4) {\bm u}\;.
 \label{eq:1}
\end{equation}
The pressure term ${\bm \nabla} p$ ensures the incompressibility of 
the flow, ${\bm \nabla } \cdot {\bm u}=0$, since in dense suspensions 
one can neglect density fluctuations. 
The coefficients $\lambda,\alpha,\beta,\Gamma_2,\Gamma_4$
are phenomenological parameters related to the properties of the 
microswimmers, the surrounding fluid and their interaction
\cite{heidenreich2016hydrodynamic,reinken2018derivation}.
In particular, large negative values of $\alpha$ correspond
to strong aligning interactions between the swimmers,
while the auto-advection parameter 
$\lambda$ is related to the motility of the 
swimmer~\cite{reinken2018derivation}.
Note that, at variance with the Navier-Stokes equation,
in the TT and TTSH models in general $\lambda \neq 1$,
therefore the model is not Galilean invariant.

For $\alpha < 0$, the Landau force $-(\alpha + \beta|{\bm u}|^2){\bm u}$
promotes the formation of collective motion with velocity 
$U=\sqrt{-\alpha/\beta}$.
This ordered state is destabilized by
the Swift-Hohenberg operator $-(\Gamma_2\nabla^2+\Gamma_4\nabla^4)$
(assuming $\Gamma_2, \Gamma_4 >0$) which generates structures
with characteristic scale $\Lambda = 2\pi \sqrt{2\Gamma_4/\Gamma_2}$
\cite{swift1977hydrodynamic,dunkel2013minimal,10.1007/978-3-319-66764-5_14}.
The scale $\Lambda$ corresponds to the typical size of the mesoscale vortices
observed in experiments \cite{wensink2012meso,dunkel2013fluid}.
Rescaling Eq.\ref{eq:1}
with the characteristic scale $\Lambda$ and velocity $U$,
the three independent dimensionless parameters of the model are
$\lambda$, $\alpha \Lambda /U$ and $\Gamma_2/\Lambda U$. 

\subsection{Numerical methods}
Equation (\ref{eq:1}) is numerically integrated in two dimensions
by a standard 
pseudo-spectral method in the vorticity-velocity formulation
with a $1/2$ dealiasing for the cubic nonlinearity, and a 4th order 
Runge-Kutta time stepping.
Confinement in a circular domain is imposed by the penalization
method \cite{arquis1984conditions,schneider2005decaying},
which consists in modeling the region outside the domain as
a porous medium with vanishing permeability. 
To this aim, the term $-\frac{1}{\tau} \mathcal{M}(\textbf{x})
\textbf{u}$ is added to (\ref{eq:1}),
where $\tau$ is the permeability
time and the mask field $\mathcal{M} (\textbf{x})$ is
equal to $0$ and $1$ respectively inside and outside
a circular domain of radius $R$ \cite{schneider2014numerical}.
 
We performed two main sets of simulations. In the first we fix 
$\lambda=3.5$ and vary both $R=\{16, 23, 32, 63\} \Lambda$ and 
$\alpha=\{-0.25,-0.50,-0.75,-1.00,-1.25,-1.50,-1.75,-2.00 \}$, 
while in the second set we fix both
$\alpha=-1.00$ and $R=63 \Lambda$, and vary 
$\lambda=\{2.0,2.5,3.0,3.5,4.0,4.5, 5.0, 7.0\}$. 
An additional set of $269$ simulations was performed with 
$\lambda = 3.5$, $\alpha = -1.50$ and $R=16 \Lambda$ in 
order to study the transition time to the circular flocking state. 
The domains with radius $R = 63 \Lambda$, $R = \{31,23\} \Lambda$ and
$R=16 \Lambda$ are embedded in squared periodic domains of size
$L = \{160, 80, 40 \} \Lambda$ with numerical resolutions
$N=\{2048, 1024, 512 \}$ respectively. 
In all the simulations the values of the other parameters are fixed
as follows: $\beta = 0.01$, $\Gamma_2 = 2$, $\Gamma_4 = 1$.
The characteristic scale is $\Lambda = 2\pi$.
The permeability time of the penalization term is $\tau = 0.001$,
which is smaller than any dynamical time.

We identify vortices by means of the standard Okubo-Weiss parameter
\cite{okubo1970horizontal,weiss1991dynamics} 
$\mathcal{Q}=\left(\partial^2_{xy} \psi \right)^2
- \left(\partial^2_x \psi \right) \left( \partial^2_y \psi \right)$, 
where $\psi$ is the stream function 
(i.e. ${\bm u}=(\partial_y \psi,-\partial_x \psi)$ and 
$\omega=-\nabla^2 \psi$ is the vorticity).
$\mathcal{Q} <0$ corresponds to vortical regions, while 
$\mathcal{Q} > 0$ to regions dominated by shear. 
Vortices are defined as connected regions of the space where 
$\mathcal{Q} \le -\mathcal{Q}^{\star}$ and the threshold value
$\mathcal{Q}^{\star}$ is chosen as $3$ times the root mean squared (rms)
value of the $\mathcal{Q}$ field. We checked that our results do not
depend on the precise value of $\mathcal{Q}^{\star}$.
The Okubo-Weiss criterion has been already used in TTSH simulations 
for the study of its Lagrangian
properties \cite{singh2022lagrangian,kiran2022irreversiblity}.

\section{Transition towards flocking turbulence}
\label{sec3}

\begin{figure*}[t!]	
\centering
\includegraphics[width=0.39\textwidth]{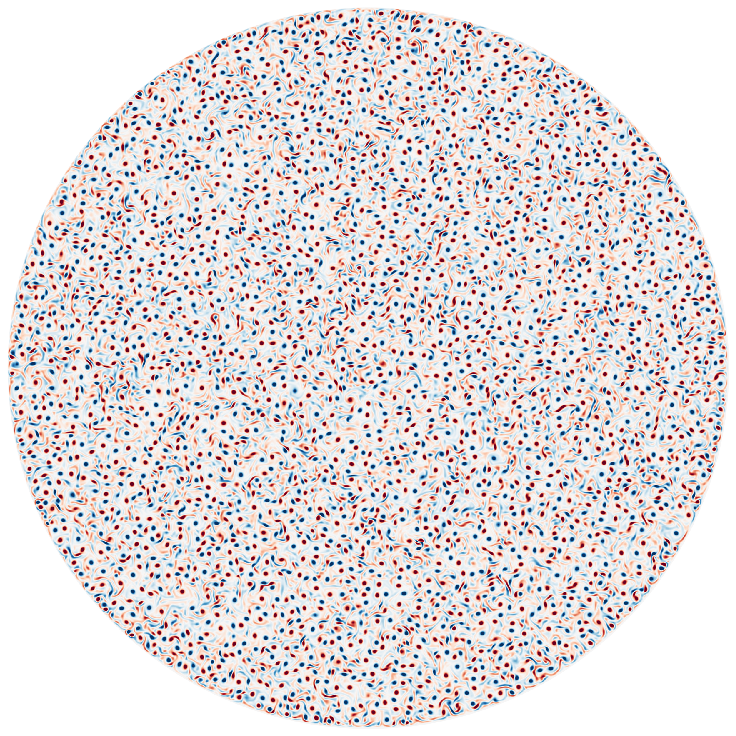}
\qquad \qquad
\includegraphics[width=0.39\textwidth]{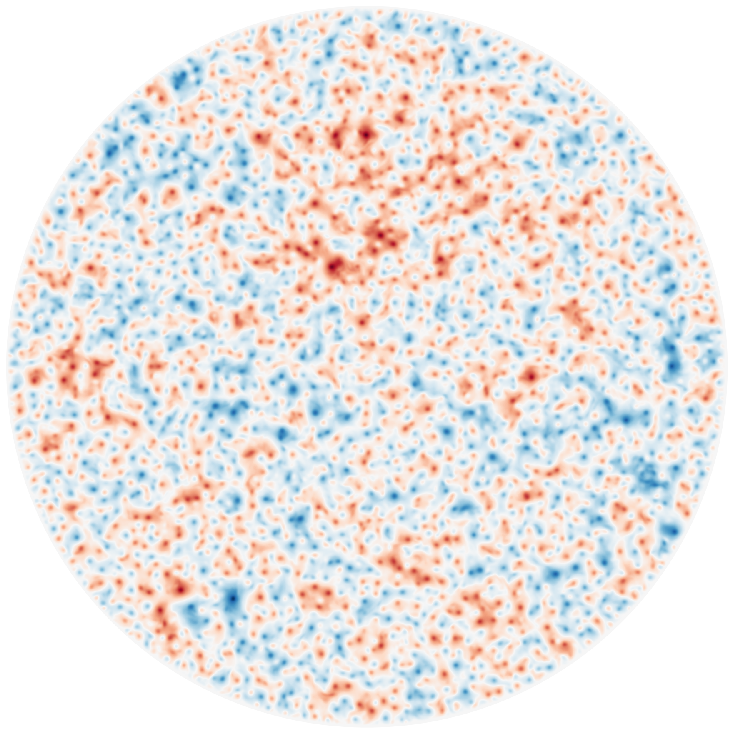}
	
\vspace{1cm}
\includegraphics[width=0.39\textwidth]{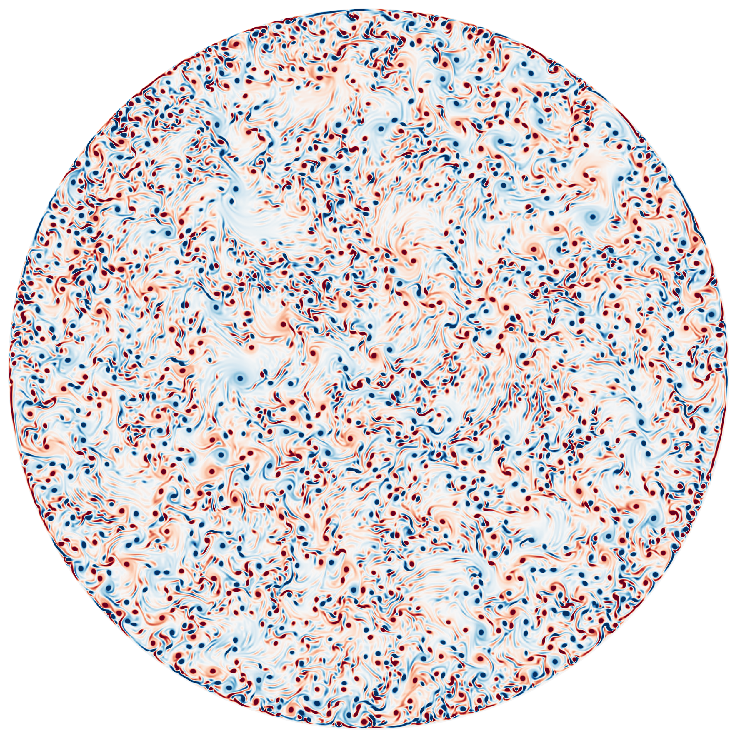}
\qquad \qquad
\includegraphics[width=0.39\textwidth]{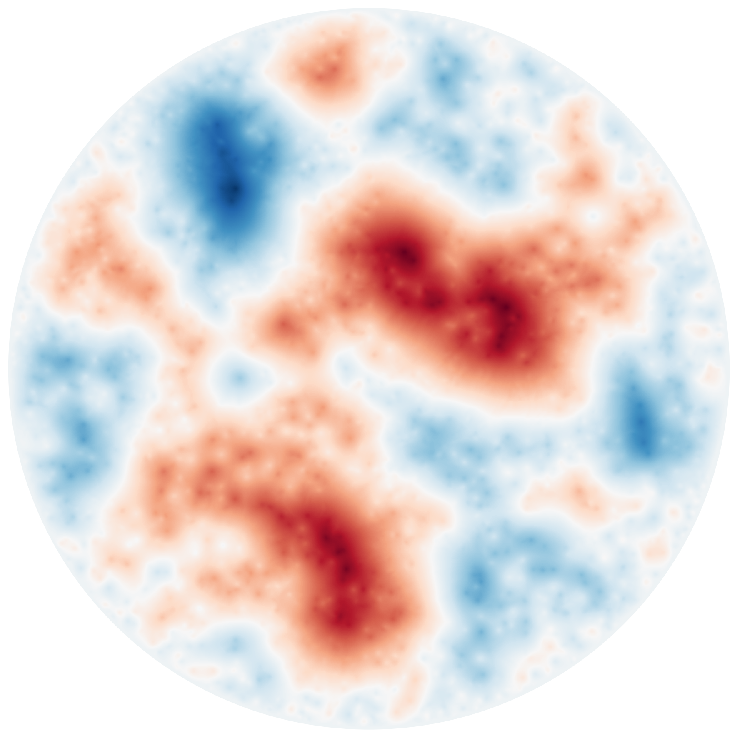}
	
\vspace{1cm}
\includegraphics[width=0.39\textwidth]{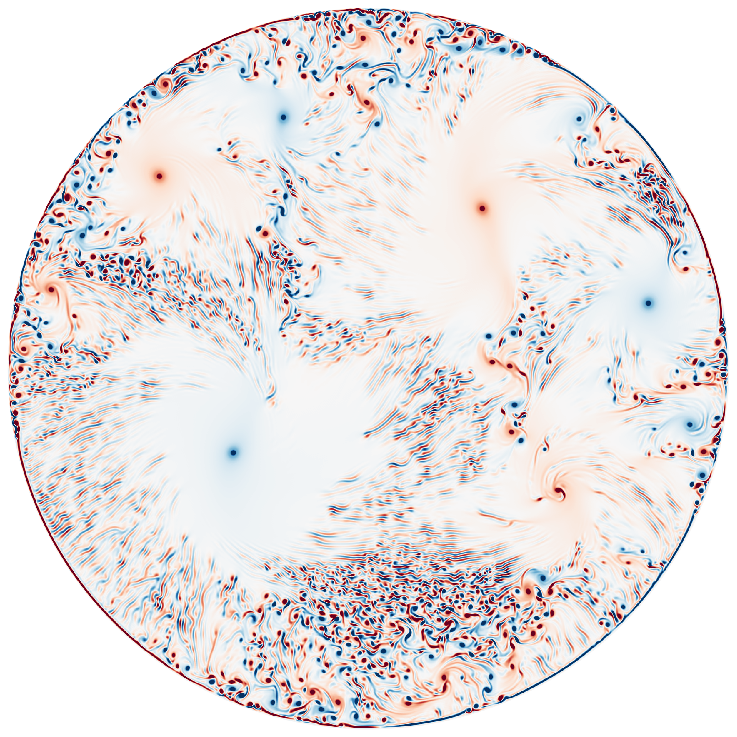}
\qquad \qquad
\includegraphics[width=0.39\textwidth]{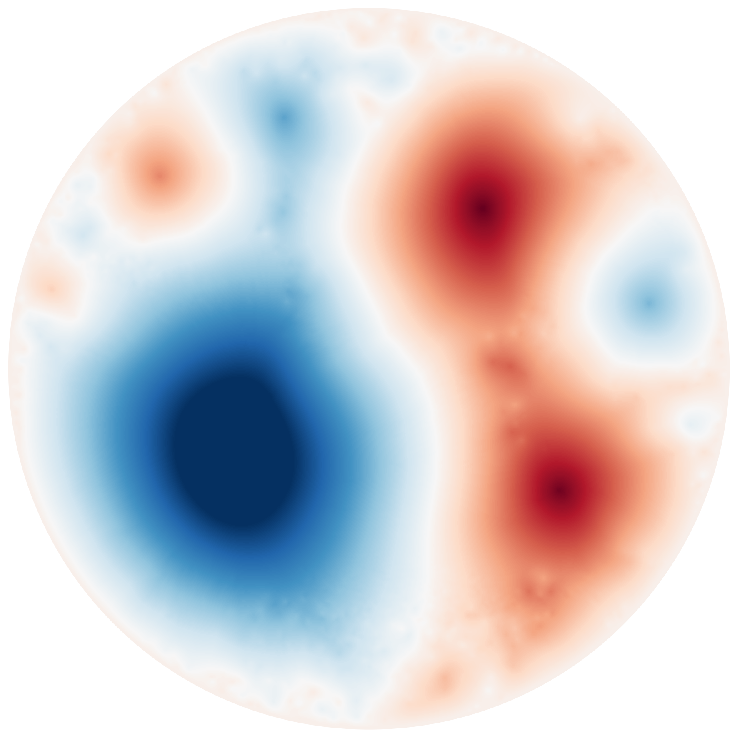}
\caption{Vorticity fields $\omega$ (left)
and stream function $\psi$ (right),
in the stationary regimes of the simulations with
$\alpha = -0.25$ (top),
$\alpha = -1.00$ (center) and 
$\alpha = -1.75$ (bottom). 
Here $\lambda=3.5$ and $R=63 \Lambda$. 
}
\label{fig1}
\end{figure*}

\begin{figure*}[t]
\centering
\includegraphics[width=0.325\linewidth]{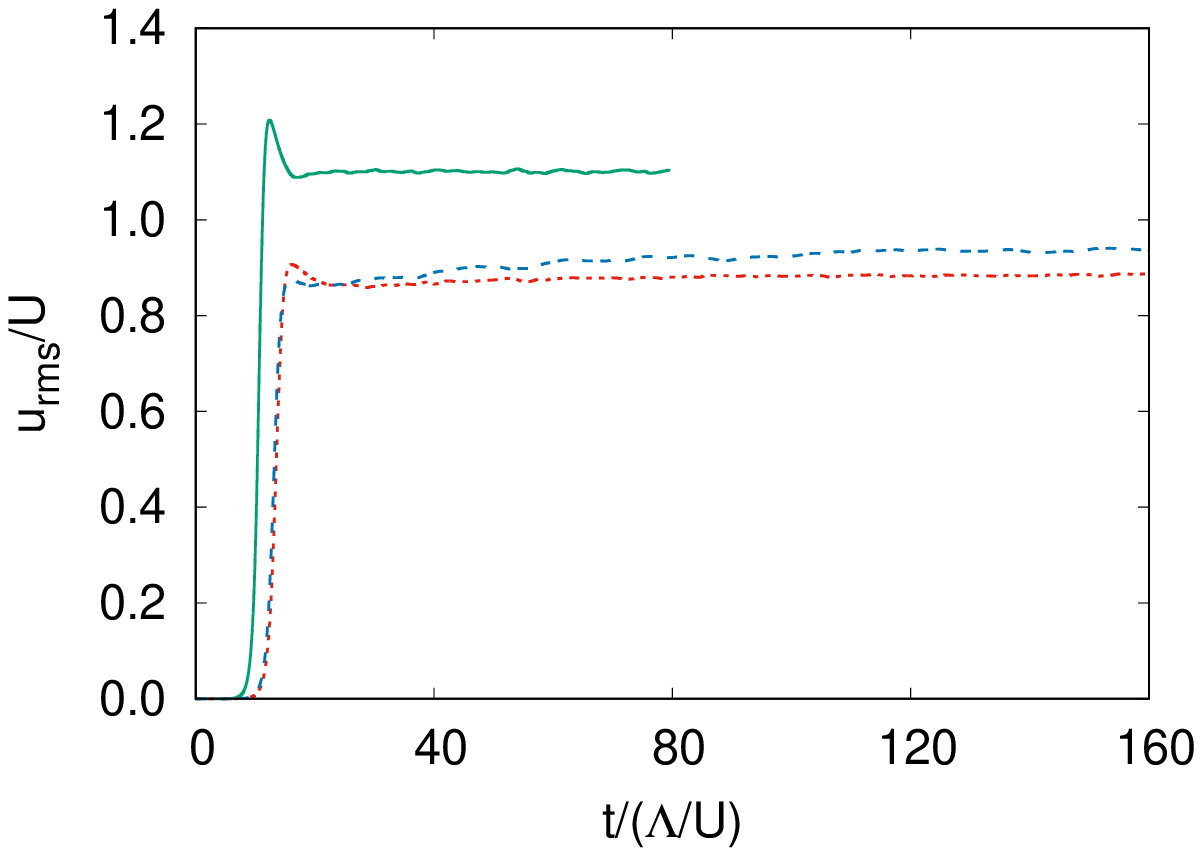}
\includegraphics[width=0.325\linewidth]{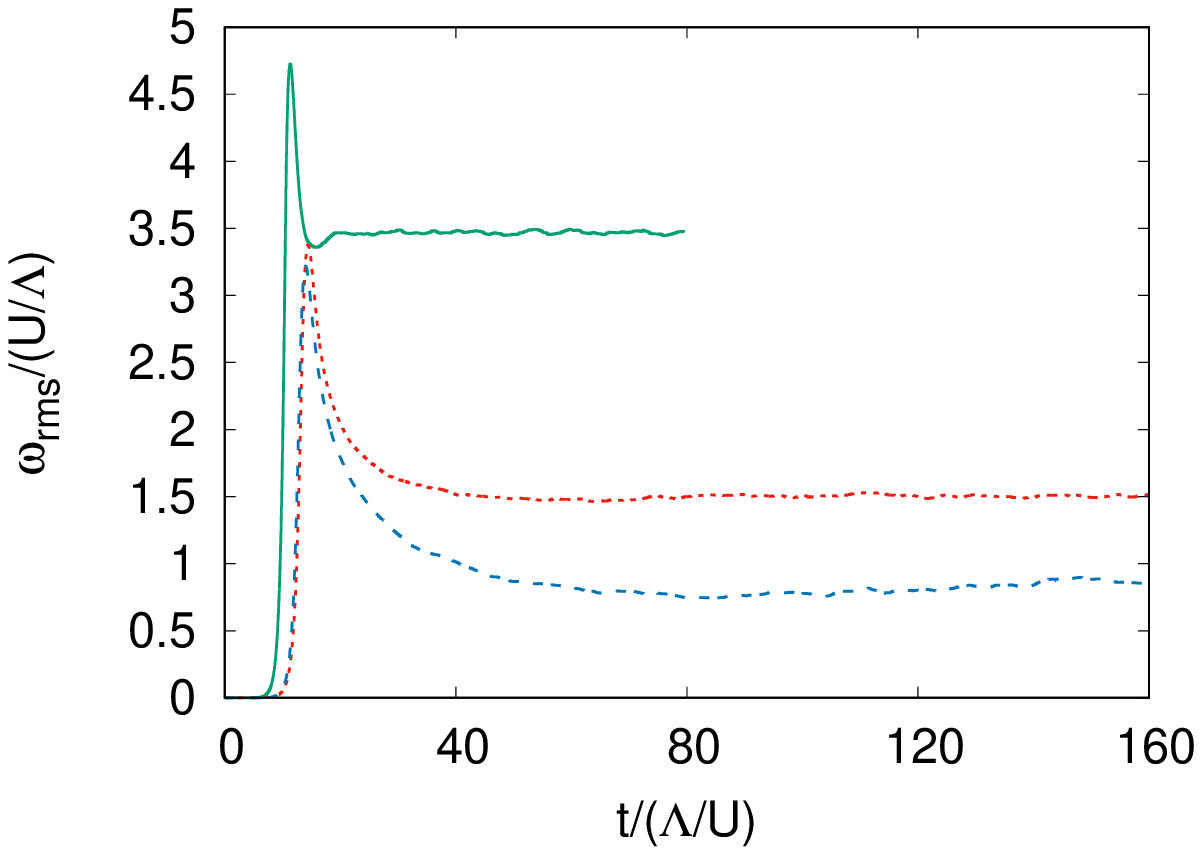}
\includegraphics[width=0.325\linewidth]{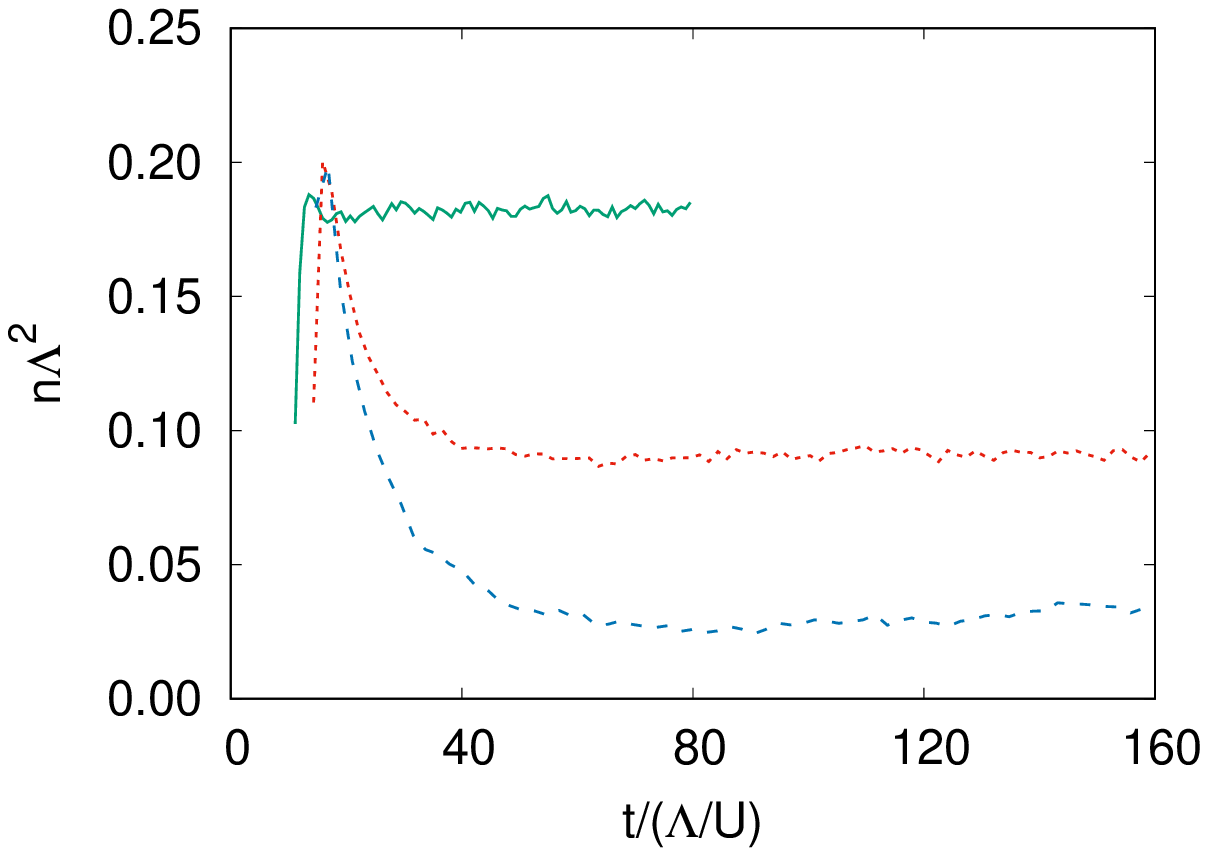}
\caption{Time evolution of the rms velocity $u_{rms}$ (left), 
rms vorticity $\omega_{rms}$ (center) and vortex density $n$ (right)
in the numerical simulations with   
$\alpha = -0.25$ (green solid line), 
$\alpha = -1.00$ (red dotted line) and
$\alpha = -1.75$ (blue dashed line). 
Here $\lambda=3.5$ and $R=63 \Lambda$. 
}
\label{fig2}
\end{figure*}

\begin{figure*}[t]
\centering
\includegraphics[width=0.325\linewidth]{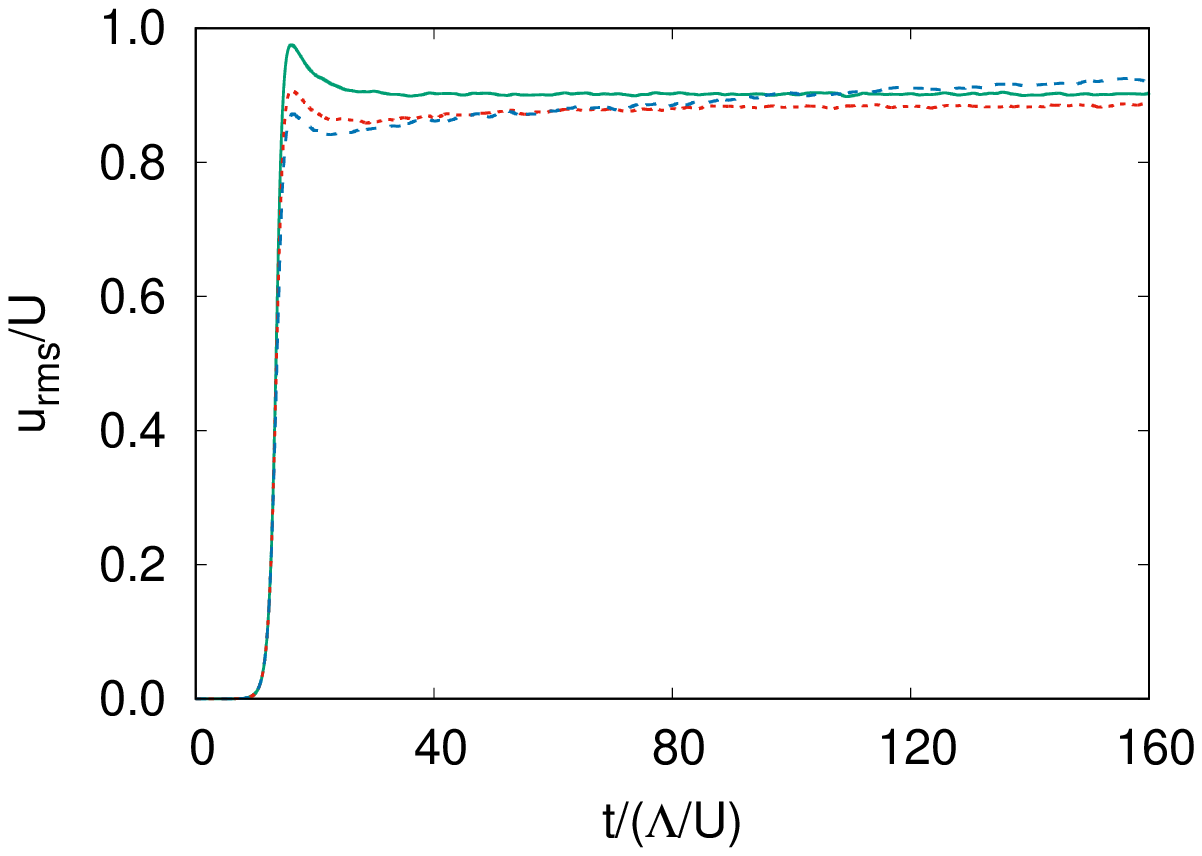}
\includegraphics[width=0.325\linewidth]{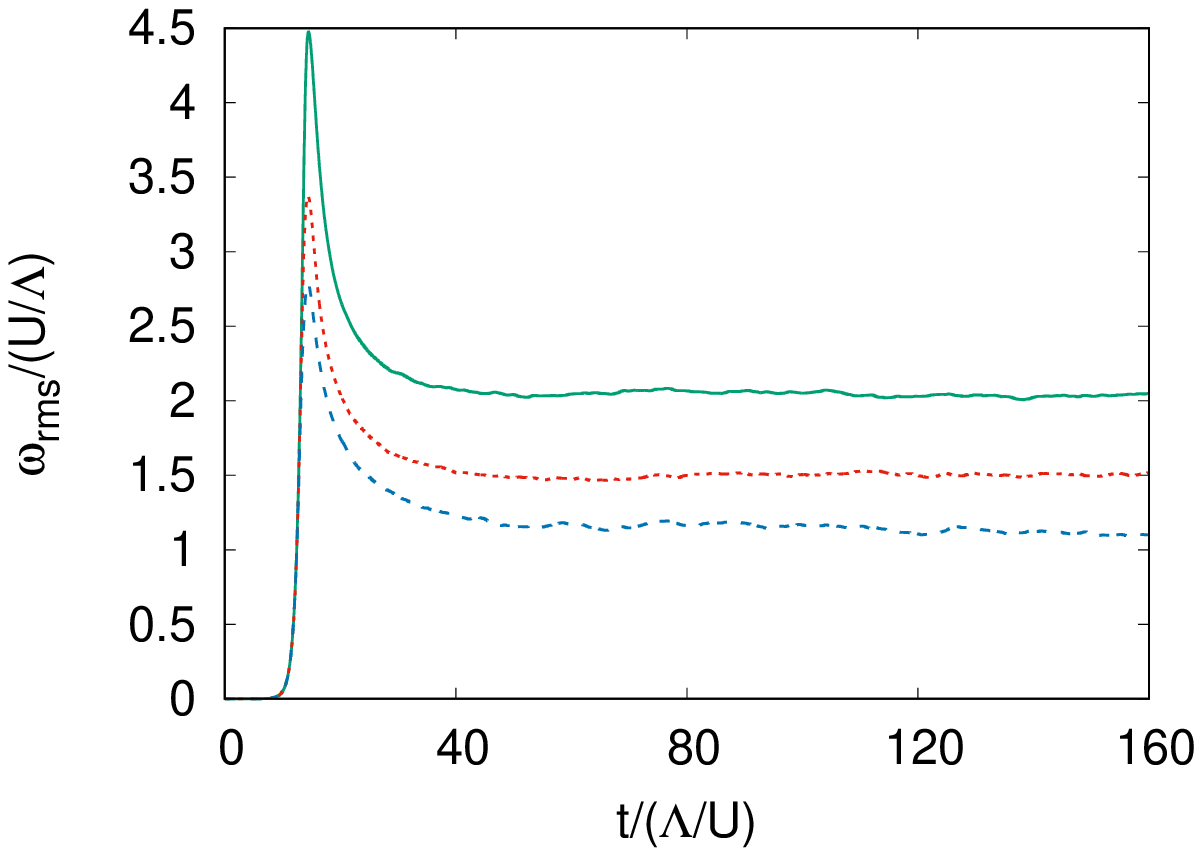}
\includegraphics[width=0.325\linewidth]{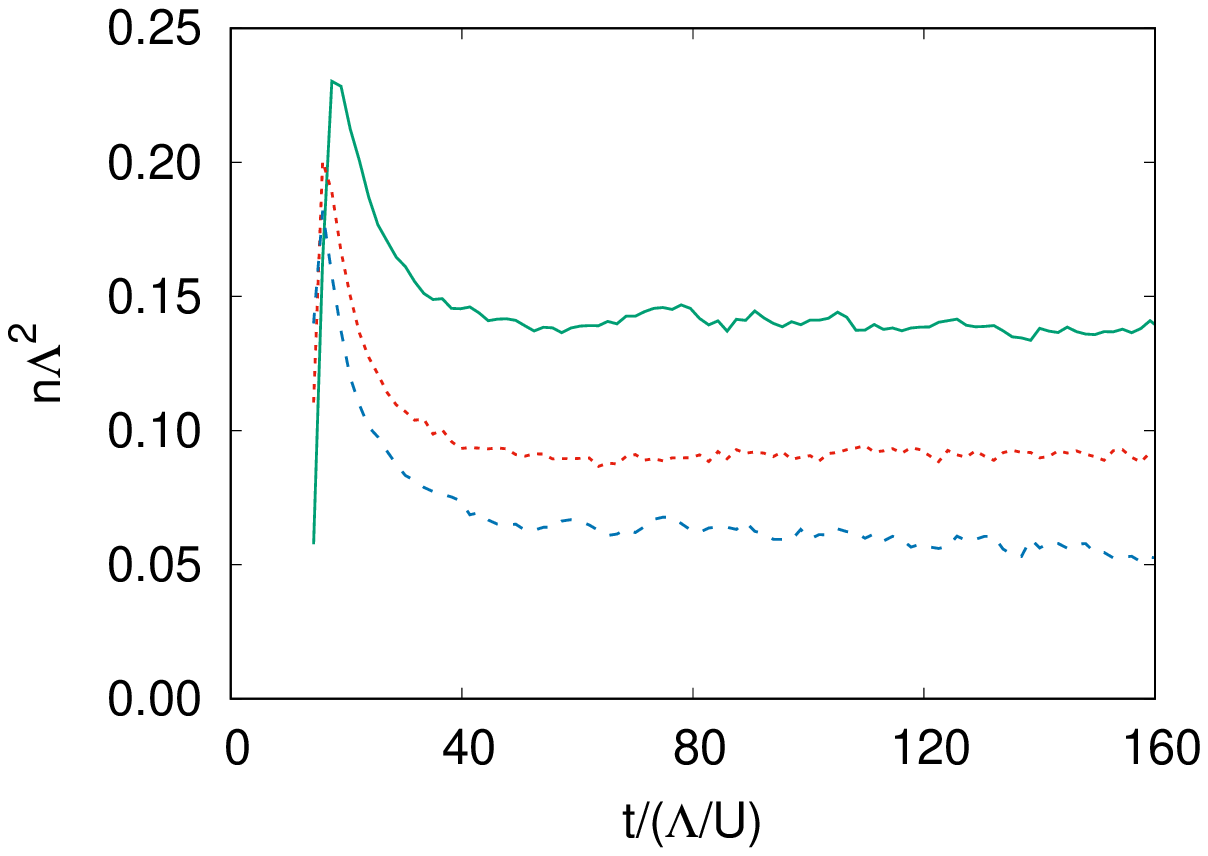}
\caption{Time evolution of the rms velocity $u_{rms}$ (left), 
rms vorticity $\omega_{rms}$ (center) and vortex density $n$ (right)
in the numerical simulations with
$\lambda=2.0$ (green solid line), 
$\lambda=3.5$ (red dotted line) and
$\lambda=5.0$ (blue dashed line). 
Here $\alpha=-1.00$ and $R=63 \Lambda$.
}
\label{fig3}
\end{figure*}

In order to study the effect of the intensity
of the aligning potential on the dynamics of the system,
we performed a first set of simulations
varying the strength of the Toner-Tu term in the range  
from $\alpha=-0.25$ to $\alpha=-2$.
In all the simulations, the initial condition is a zero velocity field
with superimposed a small random perturbation.

The early stage of the evolution of the system
is driven by the linear term
$\mathcal{L} \textbf{u}=-\alpha \textbf{u} -
\Gamma_2\nabla^2 \textbf{u} - \Gamma_4 \nabla^4 \textbf{u}$
and it is characterized by an exponential growth of the
rms values of the velocity (and vorticity)
in agreement with the predictions of the
linear stability analysis \cite{dunkel2013minimal}.
This phase ends when the nonlinear terms become relevant. 
The cubic dumping term of the Landau force arrests the exponential growth
and the self-advection term destabilizes the stationary pattern
created by the Swift-Hohenberg term, thus inducing a
mesoscale turbulence state \cite{reinken2022optimal}.
This regime is characterized by an homogeneous,
disordered population of small vortices (see Fig.~\ref{fig1} top).
The vortices are uniformly distributed in the circular domain, with a 
high vortex number density $n$ (defined as the number of vortices 
per unit area). 

For moderate negative values of $\alpha$ ($-1 \lesssim \alpha < 0$) 
the regime of mesoscale turbulence is observed to be statistically stable
and permanent. 
This is confirmed by the temporal evolution of the rms velocity $u_{rms}$,
the rms vorticity $\omega_{rms}$,
and the vortex density $n$,
which remain stationary in time (after the initial transient)
as shown in Fig.~\ref{fig2}). 

For $\alpha = -1$ the regime of uniform mesoscale turbulence loses 
its stability. 
The system undergoes a slow evolution during which the number of 
vortices diminishes
(see Fig.~\ref{fig2} left).
The decrease of the vortex density is accompanied by a decrease
of the rms vorticity, which suggests that the average vorticity
of each individual vortex remains approximately constant in time.
At long times ($t > 50 \Lambda/U$),
the system achieves an inhomogeneous, statistically steady state,
characterized by the presence of isolated vortices and 
elongate filaments called vorticity streaks~\cite{mukherjee2021anomalous}
(see Fig.~\ref{fig1} center).
The flow is organized in large-scale structures, which are evident
in the stream function.

Increasing further the strength of the aligning potential
(i.e. for $\alpha \lesssim -1$) 
the system evolves toward a strongly inhomogeneous state, 
composed by few large vortices (see Fig.~\ref{fig1} bottom).
Each vortex is surrounded by a wide region of circular motion
with constant speed $U=\sqrt{-\alpha/\beta}$.
The vorticity streaks are observed in the peripheral regions of these vortices,
and they are preferentially aligned in the transverse direction
with respect to the circular motion. 
The emergence of large-scale structures in the flow is clearly visible
in the stream function. 
Local dense vortex clusters are still present between these structures
and close to the boundary of the domain.
During the evolution of the system toward this asymptotic state,
we observe a decrease of the rms vorticity and vortices density (see Fig.~\ref{fig2}),
while the rms velocity increases slowly in time. 

The formation of this state can be understood as follows. 
At large negative values of $\alpha$, the strong Landau force promotes
the development of local attempt to organize the flow in states of circular flocking.
This process occurs independently in different regions of the domain,
producing large vortices with either positive or negative sign. 
The Swift-Hohenberg operator is too weak to suppress completely
the flocking tendency of the system,
but it is still able to destabilize the peripheral regions of the vortices. 
Indeed, linear stability analysis of a global polar 
state predicts the appearance of a transverse pattern with respect to
the mean flow with wavelength $\Lambda$ \cite{dunkel2013minimal}.
The streaks observed in Fig.~\ref{fig1} (bottom) correspond to this pattern,
distorted by the advection produced by the other vortices.
Since this regime is characterized by the chaotic interaction between
the flocking vortices, we call it \textit{flocking turbulence}.

\begin{figure*}[t]
\centering
\includegraphics[width=0.325\linewidth]{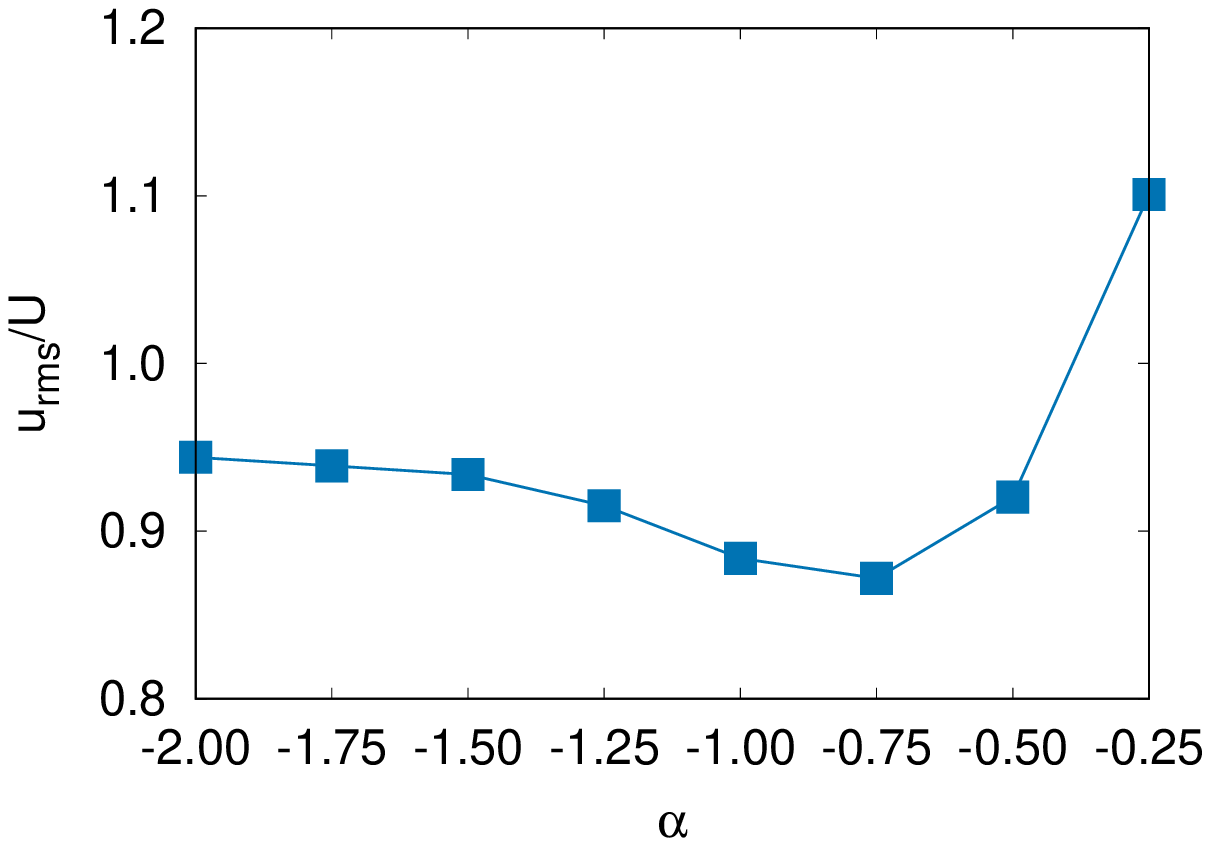}
\includegraphics[width=0.325\linewidth]{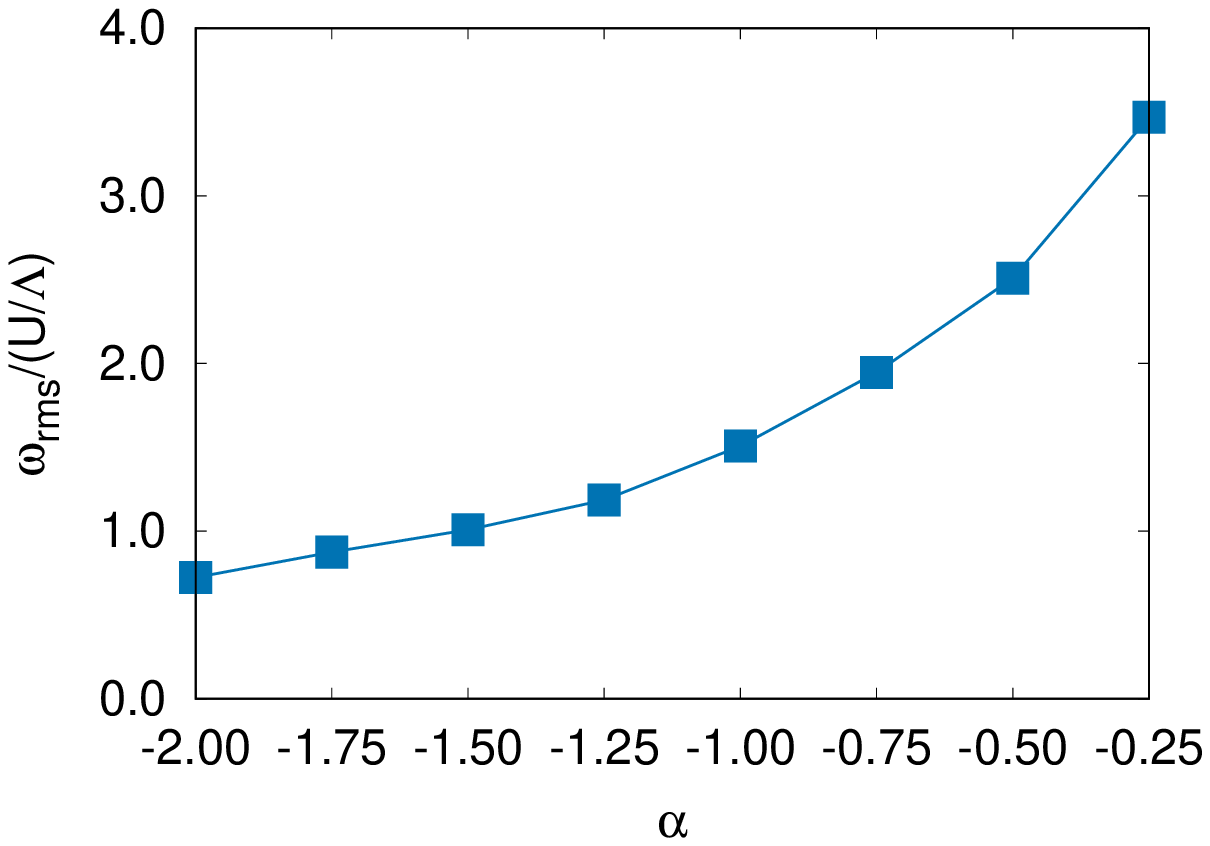}
\includegraphics[width=0.325\linewidth]{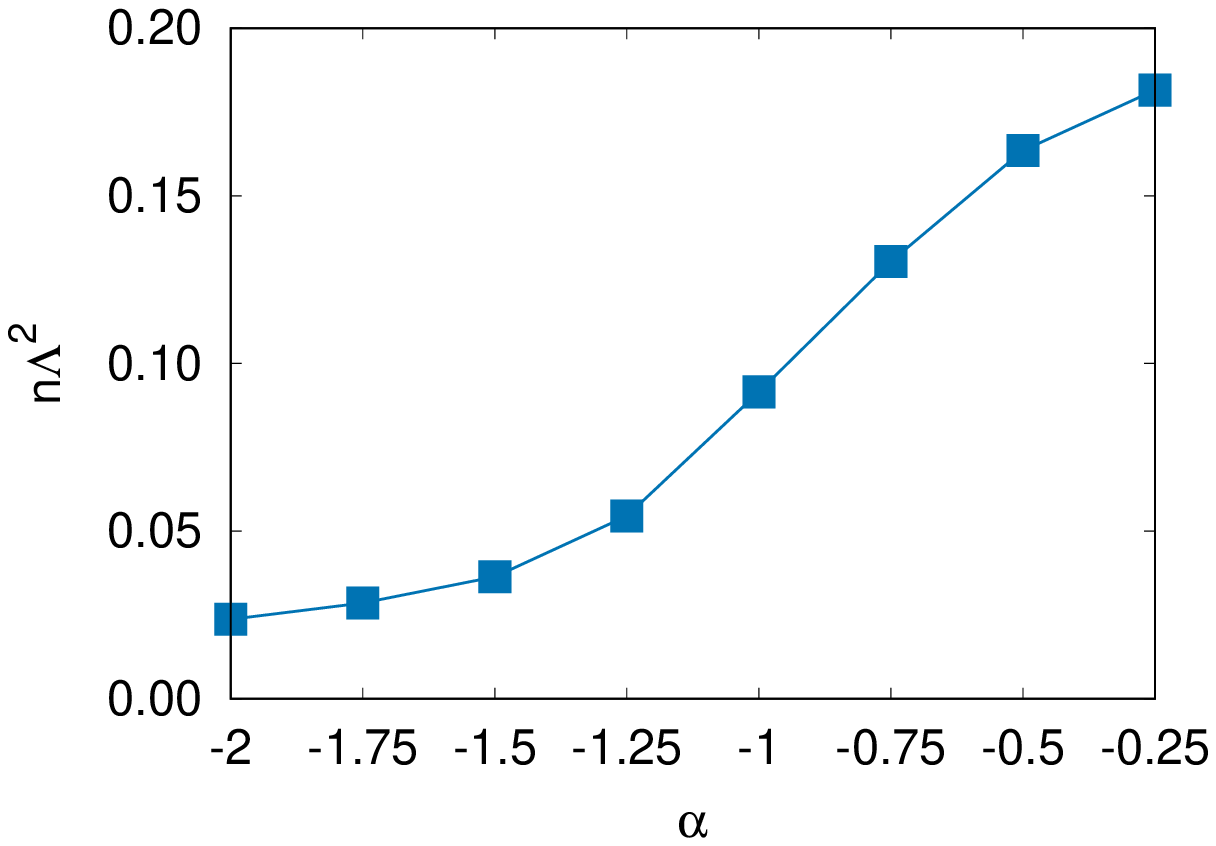}
\caption{Asymptotic values of rms velocity $u_{rms}$ (left),
rms vorticity $\omega_{rms}$ (center) and vortex density $n$ (right)
as a function of $\alpha$   
in the numerical simulations with $\lambda=3.5$ and $R=63 \Lambda$. 
}
\label{fig4}
\end{figure*}

\begin{figure*}[t]
\centering
\includegraphics[width=0.325\linewidth]{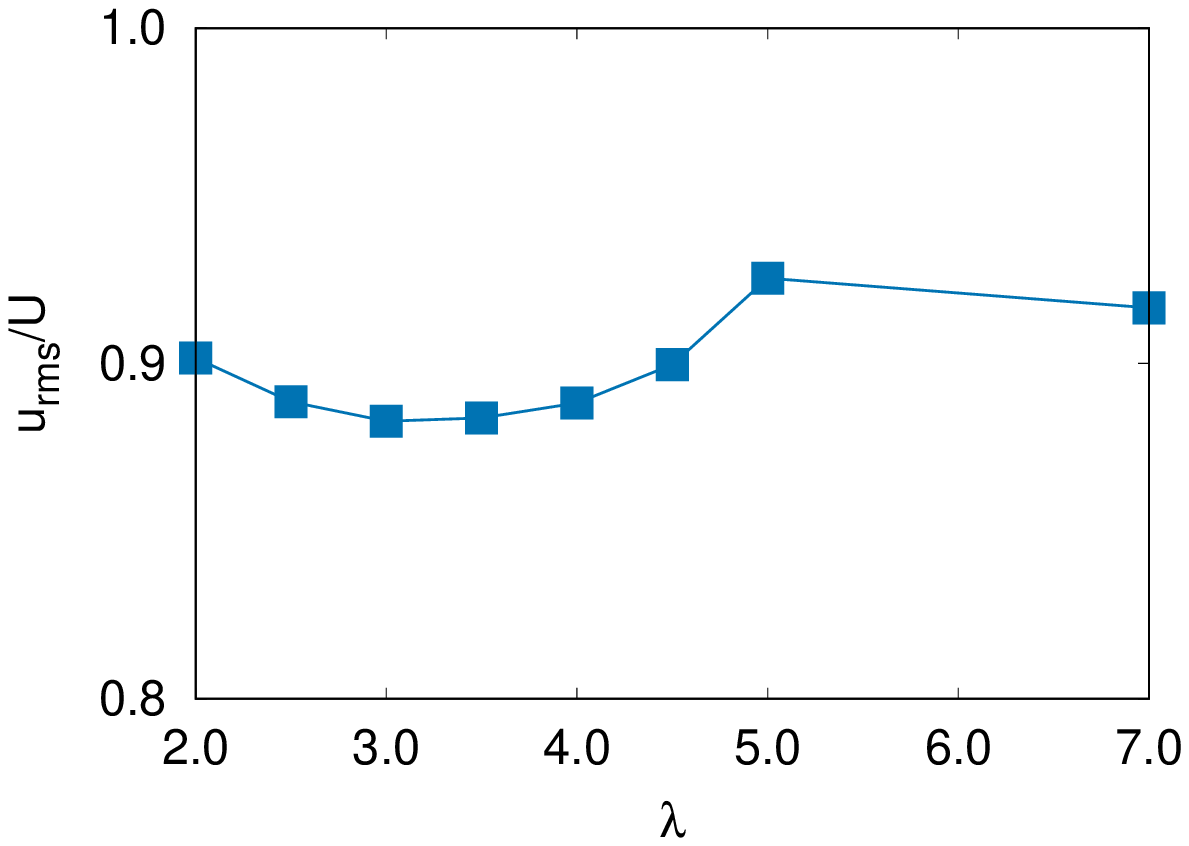}
\includegraphics[width=0.325\linewidth]{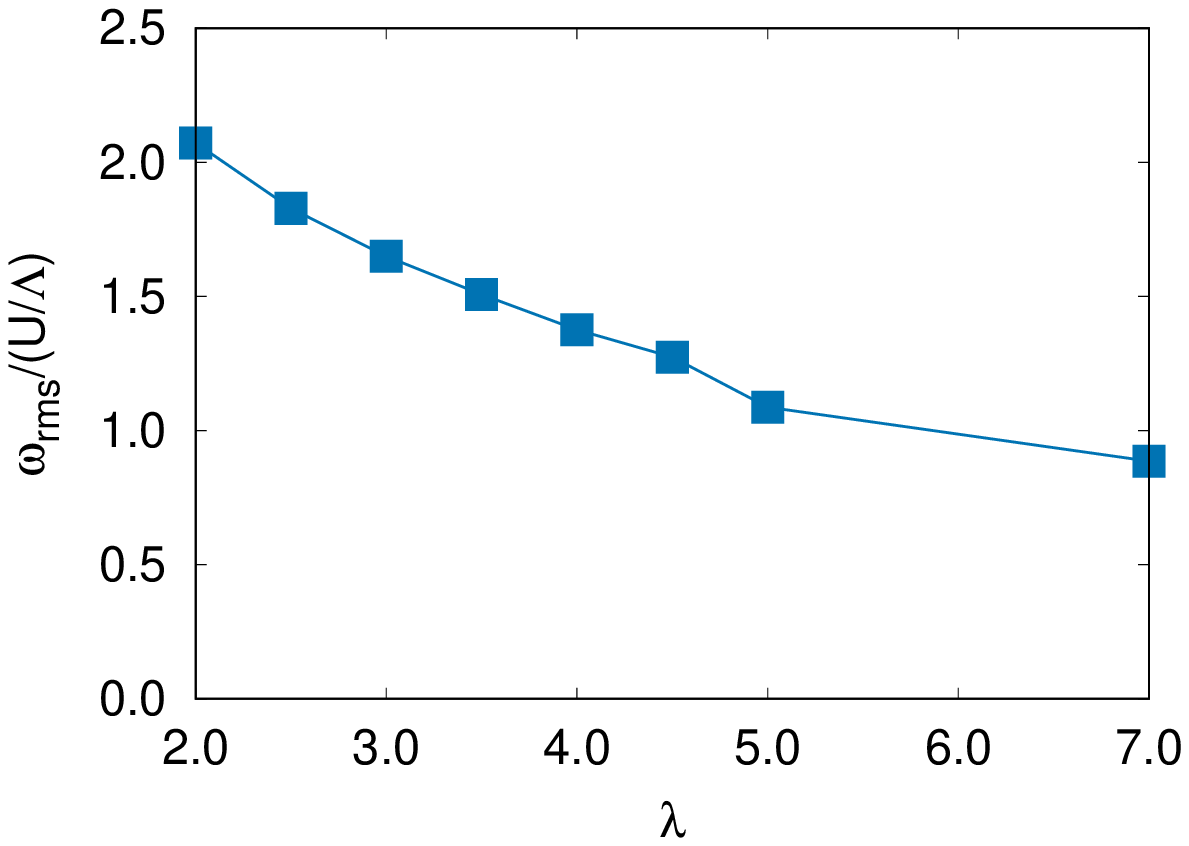}
\includegraphics[width=0.325\linewidth]{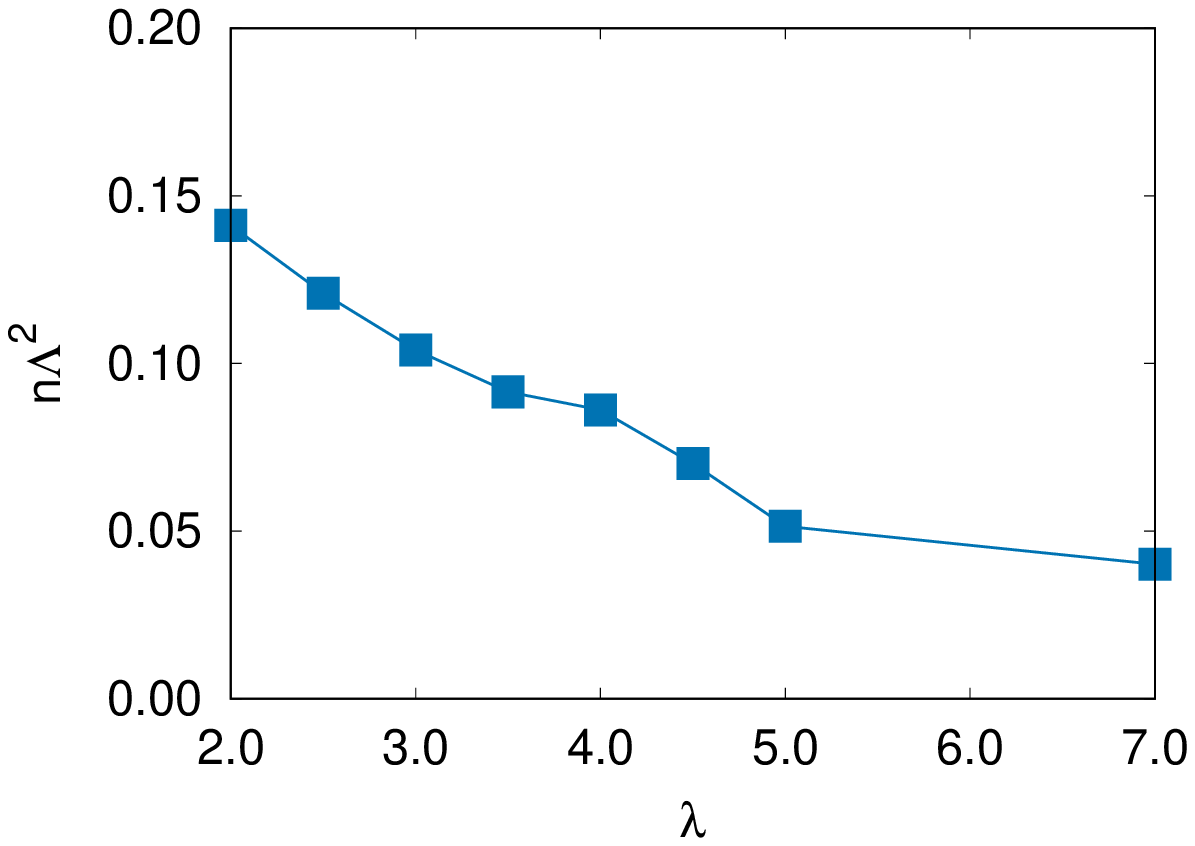}
\caption{Asymptotic values of rms velocity $u_{rms}$ (left),
rms vorticity $\omega_{rms}$ (center) and vortex density $n$ (right)
as a function of $\lambda$ 
in the numerical simulations with $\alpha=-1.00$ and $R=63 \Lambda$.
}
\label{fig5}
\end{figure*}

The non-linear self-advecting term $\lambda {\bm u} \cdot {\bm \nabla} {\bm u}$
plays a crucial role in the development of flocking turbulence. 
To address this issue, we performed a second set of simulations keeping fixed
$\alpha=-1$ and varying $\lambda$ in the range from $2$ to $7$.
The temporal evolution of the rms velocity, vorticity and vortex density
is shown in Fig.~\ref{fig3} for three different values of $\lambda$.
While the rms velocity is almost unaffected by the change of $\lambda$, 
the rms vorticity and vortex density
reach different asymptotic values which decrease at increasing $\lambda$. 
This is qualitatively similar to what observed at increasing the
intensity of $|\alpha|$ (see Fig.~\ref{fig2}).

The dependence of the asymptotic, stationary values of rms velocity,
vorticity and vortex density as a function of the parameter
$\alpha$ and $\lambda$ is shown
in Figs.~\ref{fig4} and \ref{fig5} respectively.
The transition from the two regimes of mesoscale and flocking turbulence
is evident in the dependence of the $u_{rms}$ on $\alpha$. 
In the regime of flocking turbulence, at large negative values of $\alpha$
the ratio between $u_{rms}$ and $U = \sqrt{-\alpha/\beta}$ is almost constant,
meaning that $u_{rms}$ grows proportionally to $\sqrt{|\alpha|}$.
Conversely, at small values of $\alpha$ the ratio $u_{rms}/U$ increases,
in agreement with the results of previous studies of the mesoscale
turbulence regime~\cite{PhysRevFluids.5.024302}.
We find that $u_{rms}$ is almost independent of $\lambda$.
This is consistent with the observation that the self-advection term
conserves the energy, and therefore the value of $\lambda$
is not expected to affect the energy balance. 

Both the rms vorticity and the vortex density decrease
by increasing the magnitude of $|\alpha|$,
in agreement with the qualitative observation
that the number of vortices decreases as shown in Fig.~\ref{fig1}.
A similar behavior is observed also by increasing the
strength of the self-advection: 
Larger values of $\lambda$ 
correspond to lower $\omega_{rms}$ and $n$.
The above results suggest that the transition from mesoscale turbulence
to flocking turbulence is not solely due to the increase
of the strength of the aligning potential $\alpha$, but it 
requires also a strong enough self-advection (i.e. non-linearity).

\begin{figure*}[t]
\centering
\includegraphics[width=0.49\linewidth]{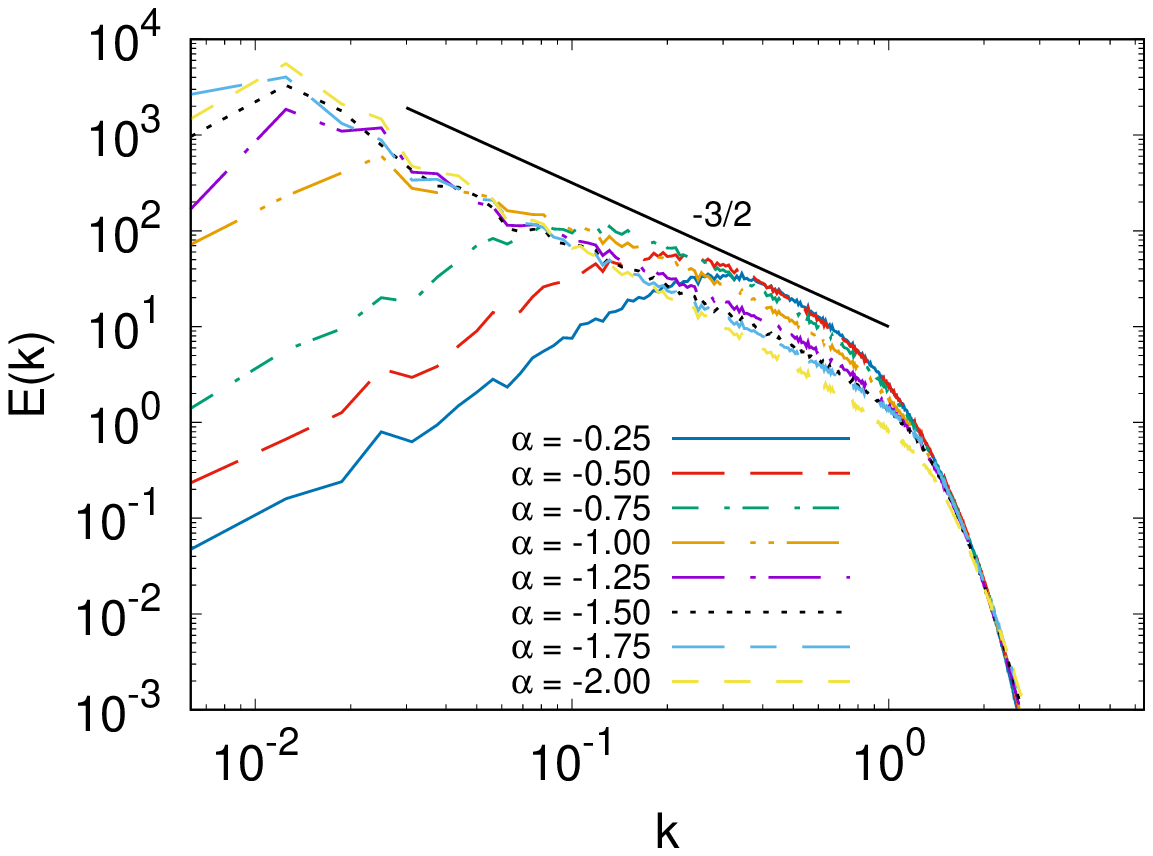}
\includegraphics[width=0.49\linewidth]{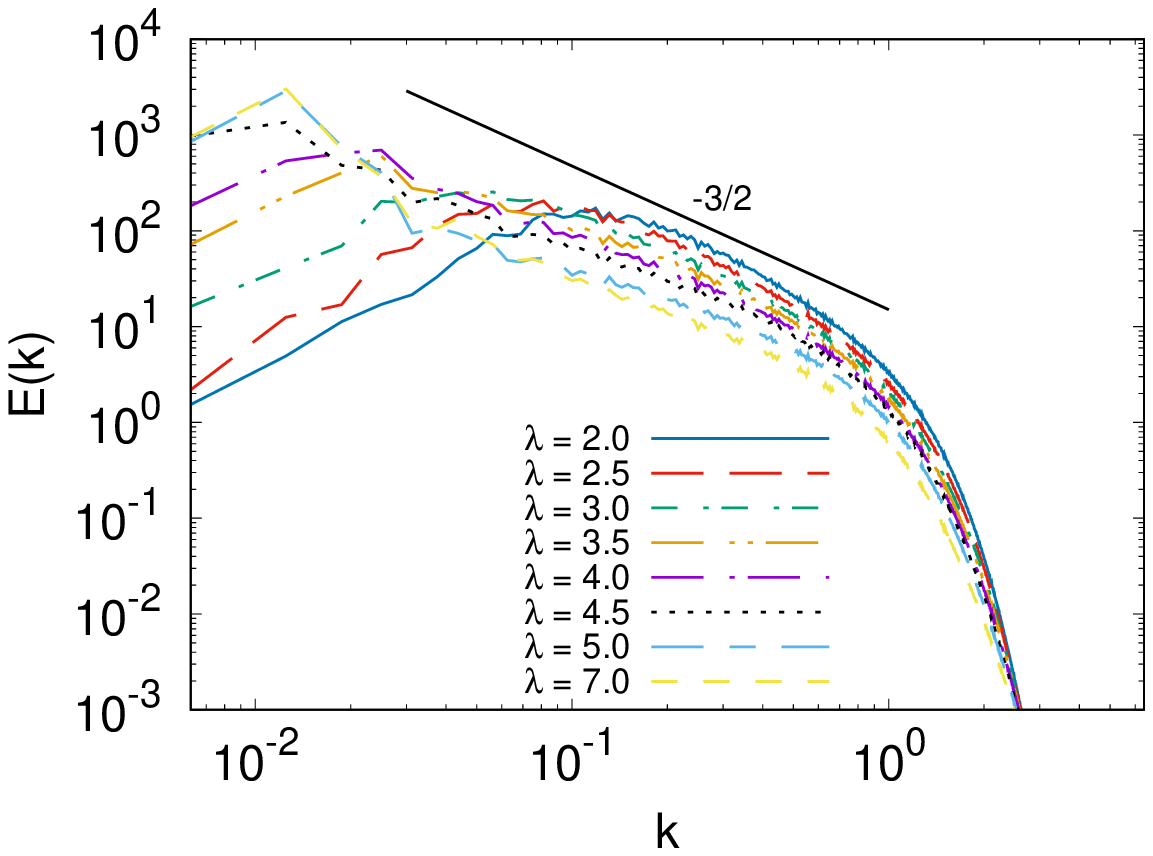}
\caption{Kinetic energy spectra,
averaged in the stationary regimes of the simulations with
fixed $\lambda =3.5$ (left) and fixed $\alpha=-1.00$ (right).
Here $R=63\Lambda$. 
}
\label{fig6}
\end{figure*}

\begin{figure*}[t]
\centering
\includegraphics[width=0.49\linewidth]{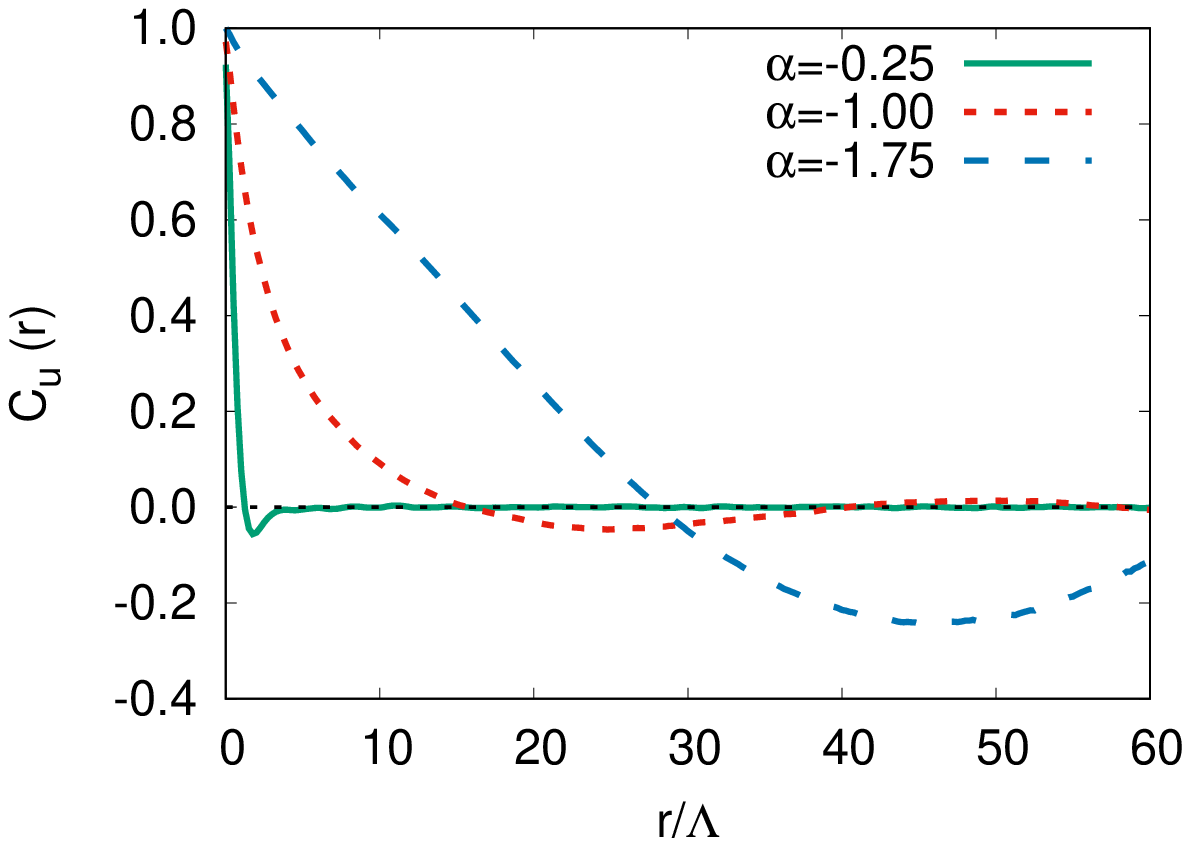}
\includegraphics[width=0.49\linewidth]{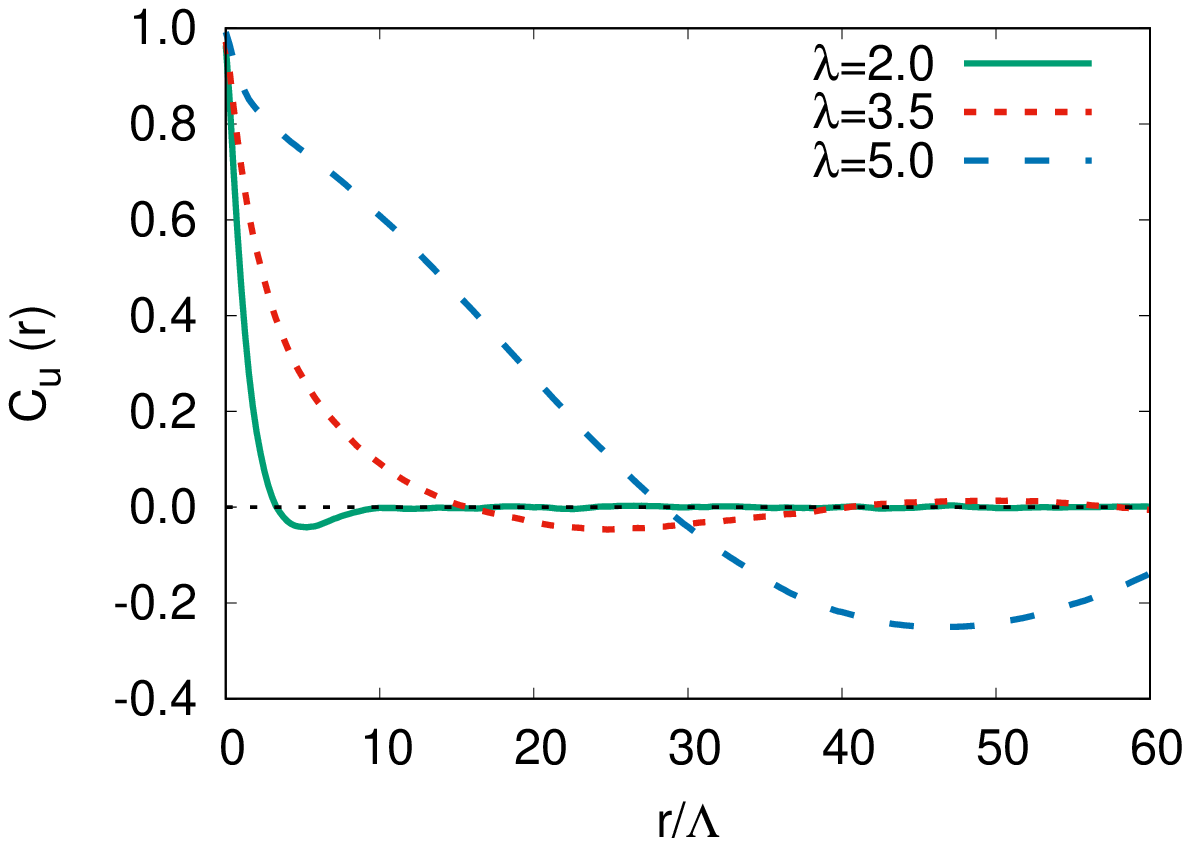}
\caption{Autocorrelations functions of the velocity field $C_{\textbf{u}}(r)$
in the stationary regimes of the simulations with
fixed $\lambda =3.5$ (left) and fixed $\alpha=-1.00$ (right).
Here $R=63\Lambda$. 
}
\label{fig7}
\end{figure*}

\begin{figure*}[t]
\centering
\includegraphics[width=0.49\linewidth]{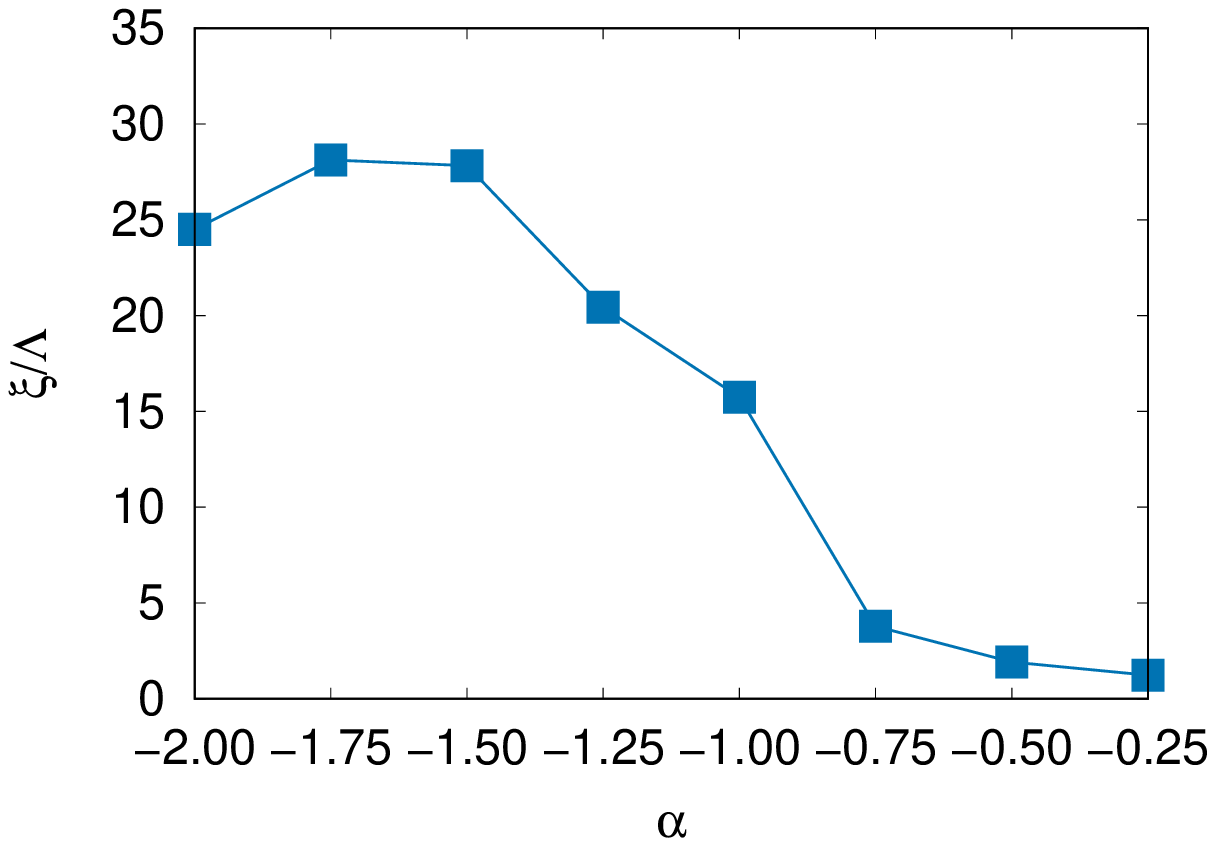}
\includegraphics[width=0.49\linewidth]{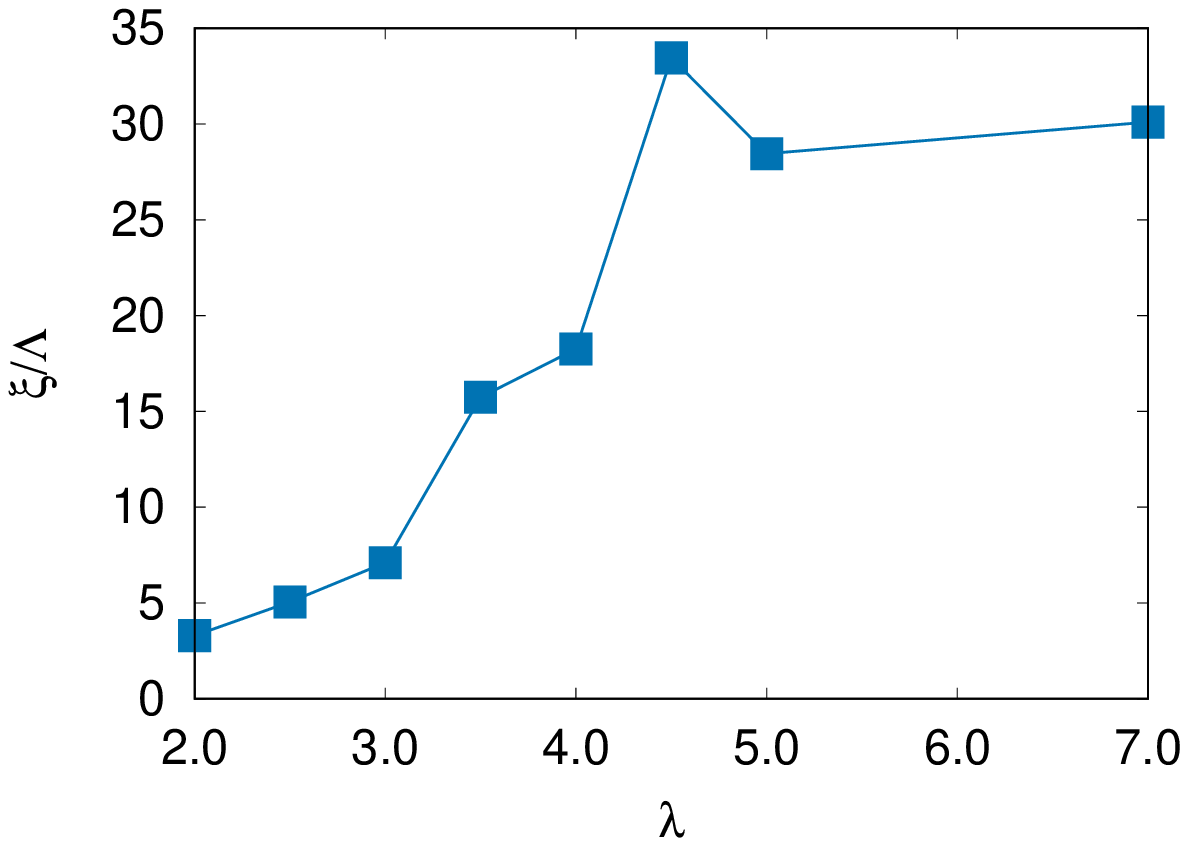}
\caption{Velocity correlation length $\xi$,
in the stationary regimes of the simulations with
fixed $\lambda =3.5$ (left) and fixed $\alpha=-1.00$ (right).
Here $R=63\Lambda$. 
}
\label{fig8}
\end{figure*}

Further insights on this transition are given by the  
distribution of kinetic energy among different spatial scales,
which is quantified by the energy spectrum $E\left( k \right)$,
shown in Fig.~\ref{fig6}.
In the mesoscale turbulence regime, 
$E\left( k \right)$ is peaked around a characteristic wavenumber 
$k_{max} \simeq 2\pi/\Lambda$, 
Increasing the energy input (i.e. increasing the magnitude of $|\alpha|$)
the peak $k_{max}$ of the spectrum shifts towards smaller wavenumbers
$k_{max} < 2\pi/\Lambda$, in agreement with previous findings
\cite{bratanov2015new,james2018vortex,PhysRevFluids.5.024302}.

In the regime of flocking turbulence (for $|\alpha| \gtrsim 1$),
we observe a qualitative change in the spectrum.
The energy spectrum develops a power-law behavior
$E(k) \sim k^{-\delta}$ at intermediate wavenumbers
$k_{max} \ll k \ll 2\pi/\Lambda$,
with a spectral slope $\delta$
which is close to the theoretical value $\delta = 3/2$
predicted and observed in~\cite{mukherjee2022intermittency}.
At large, negative values of $\alpha$ we observe
a slight increase of the spectral slope $\delta$, which exceeds the value $3/2$.
At the same time, the wavenumber $k_{max}$ becomes almost constant and it is 
close to the smallest available wavenumber,
i.e. the inverse of the size of the circular domain.
As we will discuss in the next Section, these effects are due to the confinement. 

Interestingly, we find that the decrease of peak of the energy spectrum $k_{max}$
and the development of the intermediate power-law behavior
is observed also at increasing the parameter $\lambda$ at fixed $\alpha$.
This is a further clue that the transition from
mesoscale to flocking turbulence is determined by the interplay
between the Landau force and the self-advection term. 

The growth of the integral scale of the flow, signaled by the reduction of $k_{max}$,
can be quantified by the analysis of the autocorrelation functions
of the velocity field
\begin{equation}
  C_{\bm{u}}(r) =
  \frac{ \langle \bm{u} (\bm{x}) \cdot \bm{u}( \bm{x}')\rangle}
       { \langle |\bm{u}(\bm{x})|^2 \rangle };
       \label{eq:2}
\end{equation}
with $r = \| \bm{x} - \bm{x}' \|$, and angular brackets indicating
average over space and time (in the stationary regime). 
We remark that correlation functions are a well established tool
for the study of flocking phenomena \cite{cavagna2018physics}.

The velocity autocorrelation function, plotted in Fig.~\ref{fig7},
displays a negative minimum which allows to define a velocity 
correlation scale $\xi$ given by the first zero crossing of $C_{\bm{u}}$.  
The dependence of $\xi$ on the parameters $\alpha$ and $\lambda$ is 
reported in Fig.~\ref{fig8}. 
Both reducing $\alpha$ at fixed $\lambda$ and increasing $\lambda$ at fixed $\alpha$ we
observe a sharp increase of $\xi$ from values comparable to $\Lambda$
to values of the order $30 \Lambda$,
which indicates the transition from the mesoscale regime to the flocking turbulence.
At large values of $\Lambda$ and large negative values of $\alpha$ we 
also observe a saturation of the correlation scale to an asymptotic value
$\xi \approx 30 \Lambda$
which is comparable with the radius of the circular domain ($R=63 \Lambda$).

\section{Role of confinement and circular flocking}
\label{sec4}
The saturation of correlation length and of the peak of the energy spectrum
$k_{max}$
suggests that the geometrical confinement of the bacterial 
turbulence influences 
significantly its dynamics.
In this section we pursue the investigation of the effects of the confinement
presenting the results of simulations of the TTSH model in circular domains
at varying the radius $R$ of the domain.

\begin{figure*}[t!]
\centering
\includegraphics[width=0.325\linewidth]{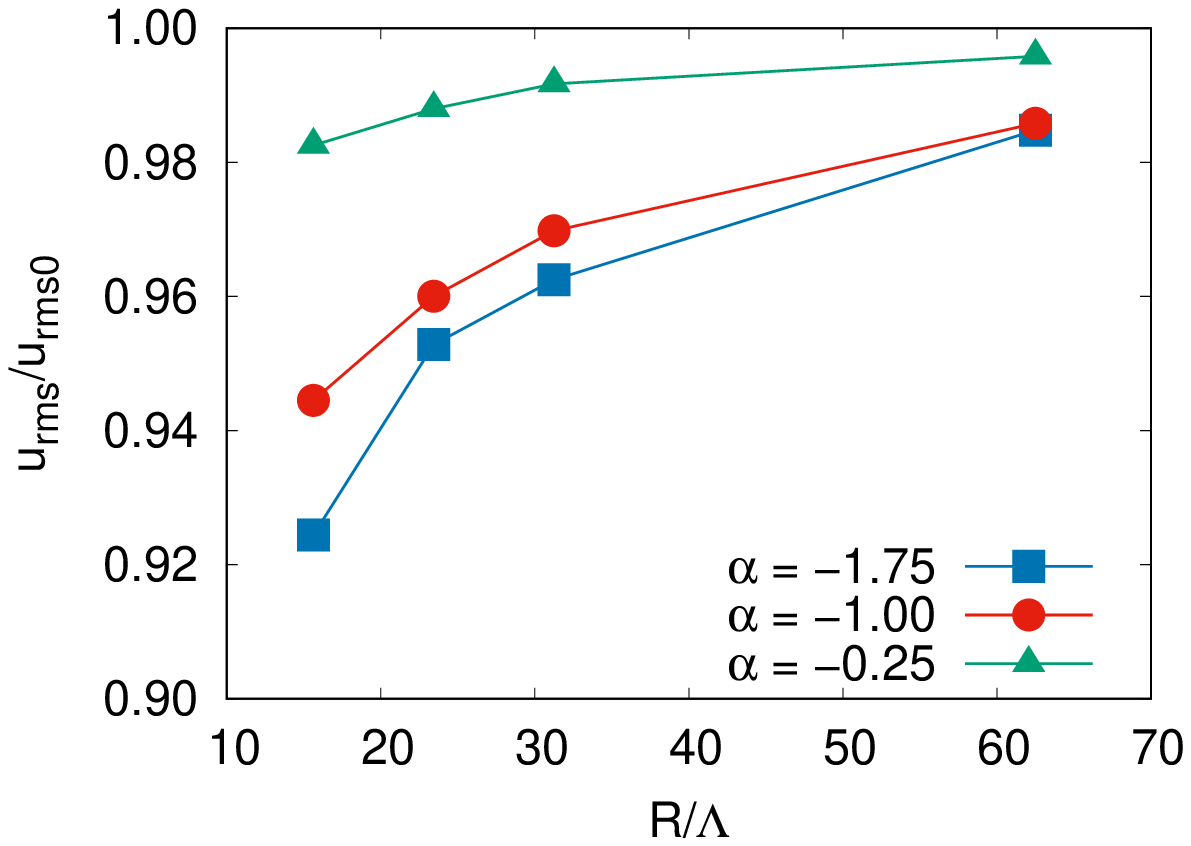}
\includegraphics[width=0.325\linewidth]{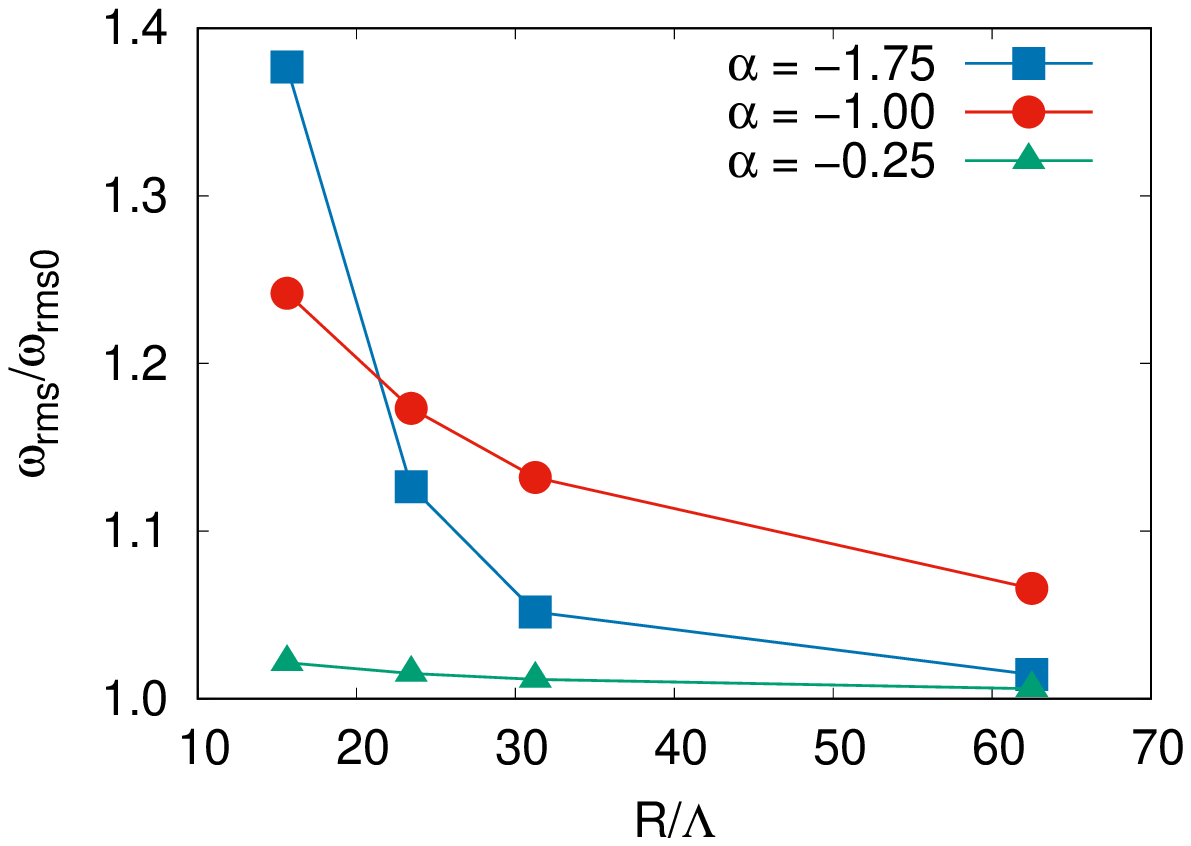}
\includegraphics[width=0.325\linewidth]{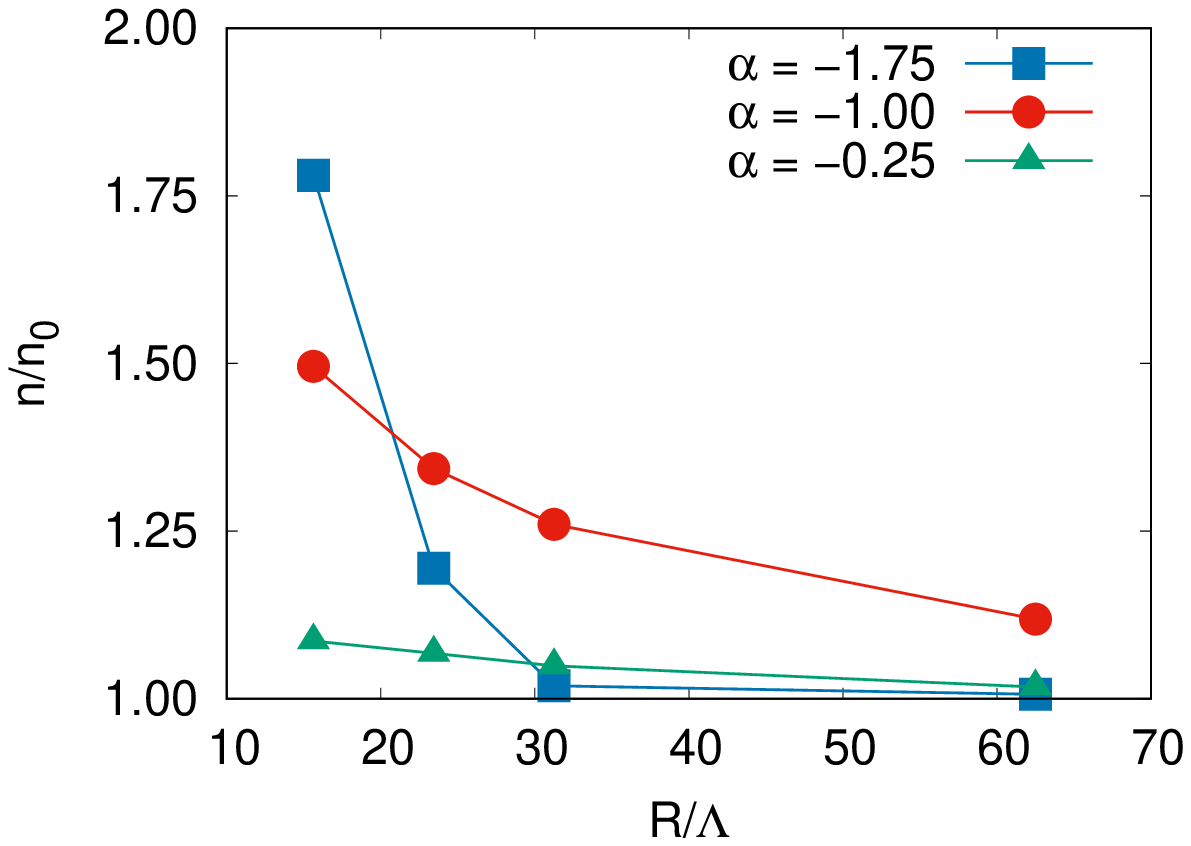}
\caption{Asymptotic values of the rms velocity $u_{rms}$ (left),
rms vorticity $\omega_{rms}$ (center) and vortex density $n$ (right)
as a function of the confinement radius $R$. Here $\lambda=3.5$.
}
\label{fig9}
\end{figure*}

In Fig.~(\ref{fig9}) we compare the asymptotic stationary values
of the rms velocity $u_{rms}$, rms vorticity $\omega_{rms}$ and 
vortex density $n$
as a function of the domain size $R$ for three sets of simulations
in the regime of mesoscale turbulence ($\alpha = -0.25$),
in the transition regime ($\alpha = -1.00$)
and in the regime of flocking turbulence ($\alpha=-1.75$).
The parameter $\lambda=3.5$ is fixed for all the simulations.
The asymptotic values presented in Fig.~(\ref{fig9})
are normalized with the corresponding values
$u_{rms 0}$, $\omega_{rms 0}$ and $n_0$
obtained in another set of simulations with identical parameters,
performed in a large square domain with size $L = 160 \Lambda$ and periodic BCs,
which is the typical setup for the numerical studies of the TTSH model.

The effects of the confinement are qualitatively similar for all the regimes:
Increasing the confinement, i.e. reducing $R$, we observe a decrease
of the asymptotic values of $u_{rms}$ and an increase of $\omega_{rms} $ and $n$.
This effects can be ascribed to the interactions of the flow with the no-slip boundary.
The friction with the boundary dissipates part of the energy, thus reducing $u_{rms}$.
Close to the boundaries, the energy dissipation is accompanied by the production
of small vortices, which causes an increase of $\omega_{rms}$ and of the total
number of vortices. These effects are stronger for the cases with smaller radius $R$,
because of the larger ratio between the perimeter and the area of the domain.

Nonetheless, we observe significant quantitative differences.
In the case of mesoscale turbulence ($\alpha =-0.25$)
the values of $u_{rms}$, $\omega_{rms}$ and $n$ varies weakly with $R$
and they remains close to those of the simulations with periodic BCs.
Conversely, the values obtained in the flocking turbulence regime
displays a strong dependence on $R$. 
The weak influence of the confinement on the mesoscale turbulence
can be explained by the observation that in this regime the correlation length $\xi$
of the velocity field is approximately one order of magnitude smaller than $R$. 
Therefore the effects of the confinement are restricted to a small region 
close to the boundary. 
In the intermediate case ($\alpha =-1$) the correlation length $\xi$
is larger than in the mesoscale turbulence (see Fig.\ref{fig8})
and the effects of the confinement are stronger. 
In the flocking turbulence regime ($\alpha =-1.75$) the values of
$u_{rms}$, $\omega_{rms}$ and $n$ change rapidly when the radius $R$
becomes smaller than the correlation length $\xi \approx 30 \Lambda$. 

\begin{figure}[th!]
\includegraphics[width=0.5\textwidth]{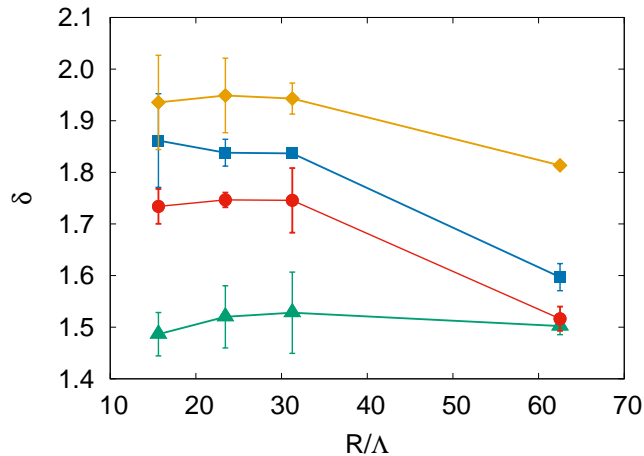} 
\caption{
Spectral exponent $\delta$ as a function of confinement radius $R$,
in the simulations with $\lambda=3.5$ and 
$\alpha=-2.00$ (orange diamonds),
$\alpha=-1.75$ (blue squares),
$\alpha=-1.50$, (red circles) and
$\alpha=-1.25$ (green triangles).
}
\label{fig10}
\end{figure}

The effect of the confinement is evident also in the energy spectra.
In Fig.~\ref{fig10} we compare the spectral slope $\delta$
of the energy spectrum measured in simulations with different $R$ and $\alpha$.
In the set of simulations with $\alpha = -1.25$ the slope
is almost independent on $R$
and its value is close to the theoretical prediction $3/2$
\cite{mukherjee2022intermittency}.
The independence of the spectra on $R$ is observed also for $\alpha > -1.25$ (not shown),
which confirms that the effects of the confinement
on the regime of mesoscale turbulence are weak.

In the regime of flocking turbulence the spectral slope $\delta$
varies significantly with $R$ and $\alpha$. 
Decreasing the radius $R$ we find that $\delta$ grows up to an asymptotic
value which increases with $|\alpha|$. 
We argue that the steepening of the energy spectrum due to the confinement
can be related to the process of spectral energy condensation,
i.e., of the accumulation of energy in the lowest mode accessible to the system,
whose wavelength is comparable to the size of the domain.
This explains the discrepancy between the slope of the energy spectra observed
in our simulations and the results reported in \cite{mukherjee2022intermittency}.
The trend of the values of $\delta$ at increasing $R$ suggests the conjecture
that the spectral slope attains an universal value $\delta =3/2$ 
in the limit of unconfined, infinite domain.

Recent studies \cite{puggioni2022giant} showed
that in the case of strong confinement (i.e., small $R$) 
the regime of flocking turbulence becomes metastable.
After a long transient, it displays a sudden transition
towards an ordered state of circular flocking,
which corresponds to a single giant vortex which spans the whole domain.
The giant vortex is surrounded by an ordered pattern of vorticity streaks,
aligned in the radial direction. 

A convenient indicator of the transition to the giant vortex regime
is provided by the vortex order parameter 
\cite{wioland2013confinement,lushi2014fluid}
\begin{equation}
  \Phi = \frac{\langle u_{\varphi}\rangle/ \langle | \bm{u} | \rangle - 2/\pi}{1- 2/\pi},
  \label{eq:3}
\end{equation}
which is defined in terms of the radial and angular components of the velocity field
$\bm{u} = u_r \hat{\bm{r}} + u_{\varphi} \hat{\bm \varphi}$.
In the case of randomly oriented velocity field one has $\Phi=0$,
while $\Phi=1$ for a velocity field fully oriented in the angular direction.

\begin{figure}[th!]
\includegraphics[width=0.5\textwidth]{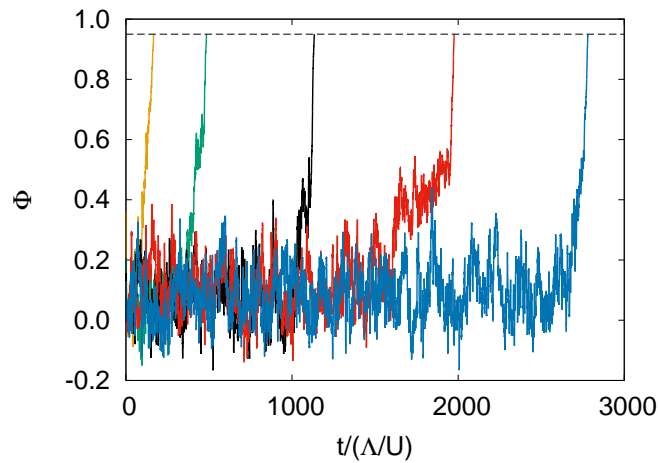} 
\caption{Time evolution of the vortex order parameter $\Phi$
in an ensemble of simulations with
$\lambda=3.5$, $\alpha = -1.50$ and $R=16 \Lambda$,
with different initial conditions.
}
\label{fig11}
\end{figure}

\begin{figure}[th!]
\includegraphics[width=0.5\textwidth]{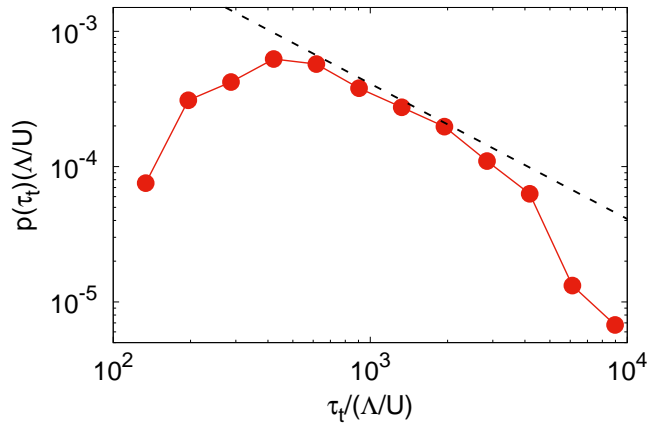} 
\caption{Probability density function $p\left( \tau_t \right)$
of the transition time $\tau_t$ in the numerical simulations with 
$R=16\Lambda$, $\alpha=-1.50$ and $\lambda=3.5$.
The black dashed line represents the slope $\tau_t^{-1}$.
}
\label{fig12}
\end{figure}

In Fig.~\ref{fig11} we show the time evolution of $\Phi$ of an
ensemble of simulations with $\alpha=-1.50$, $\lambda=3.5$ and $R=16 \Lambda$, 
starting from different initial random perturbations of the null velocity field.
Initially, the order parameter fluctuates around values close to 0,
which is due to the disordered flow in the regime of flocking turbulence.
After a long transient time $\tau_t$ we observe
a rapid increase of $\Phi$ to a value close to $1$,
which signals the transition to an ordered state.

Although the simulations are performed with identical parameters, 
the transition time exhibits an extremely large variability on the 
initial condition, and this reflects the stochastic nature of the transition.
The distinctive feature of the regime of flocking turbulence is the presence
of many large-scale vortices (see Fig~\ref{fig1} bottom),
which represent independent attempts to organize the flow in states of local flocking.
In a confined domains, the competition between different vortices
ends when one of them prevails over the others.
Because of the chaotic dynamics of the vortices,
the time at which this phenomenon occurs is unpredictable. 

In order to quantify the variability of the transition times $\tau_t$,
we show in Fig.~\ref{fig12} the probability distribution function (pdf)
of $\tau_t$, obtained from an ensemble of 269 simulations
with $\alpha=-1.50$, $\lambda=3.5$ and $R=16 \Lambda$
and different initial perturbations.
The transition time $\tau_t$ is defined
as the time at which $\Phi$ exceeds the threshold value $\Phi_{thr} = 0.95$.
The values of $\tau_t$ span over almost two orders of magnitude,
with a maximum of the pdf at $\tau_t \sim 40 \Lambda/U$.
In the range $500 < \tau_t/(\Lambda/U) < 4000$,
the pdf displays a power-law behavior $p(\tau_t) \sim \tau_t^{-1}$
followed by a rapid decay at larger times. 

\section{Conclusions}
\label{sec5}

We presented a numerical investigation of the dynamics
of dense suspensions of microswimmers, 
described by the Toner-Tu-Swift-Hohenberg model
confined in circular domains.
We explored the parameter space of the model
by varying the intensity of the aligning Landau force $\alpha$,
the self-advection parameter $\lambda$
and the radius of the circular domain $R$. 

We find that the model displays two different regimes,
which are observed respectively at moderate 
and large negative values of $\alpha$.
At weak intensity of the Landau force, the velocity field consists of
an homogeneous dense population of small vortices
which move chaotically.
This regime is known in the literature as mesoscale turbulence.
Increasing the intensity of the aligning force
(i.e. at large negative values of $\alpha$)
we observe the emergence of the flocking turbulence regime, 
characterized by an inhomogeneous flow with 
large-scale vortices surrounded by regions of uniform circular motion,
alternated with regions of elongated vortical structures called streaks. 

The transition between the two regimes is determined by the competition
between the Landau (or Toner-Tu) force $-(\alpha + \beta |{\bm u}|^2){\bm u}$,
which promotes the self-organization of the swimmers in flocks with uniform velocity
$U = \sqrt{-\alpha/\beta}$, 
and the Swift-Hohenberg term $-(\Gamma_2 \nabla^2 + \Gamma_4 \nabla^4){\bm u}$,
which destabilizes the flocks and
causes the formation of small vortical structures with typical size
$\Lambda = 2\pi \sqrt{2\Gamma_4/\Gamma_2}$.
The regime of mesoscale turbulence occurs
if the Landau force is weaker than the Swift-Hohenberg term.
In this case, the Landau force acts simply as a forcing term
which sustains the motion of the swimmers, but they cannot develop large flocks.
Conversely, when the Landau force prevails,
we observe the spontaneous formation of local circular flocks with constant speed $U$. 
Since the velocity field ${\bm u}$  represents also the order parameter of the system,
we can interpret the formation of these coherent structures 
as a spontaneous breaking of the local rotational symmetry,
occurring independently in different regions of the domain.

Our study demonstrated that the self-advection
term $\lambda {\bm u} \cdot \nabla {\bm u}$
plays an important role in this transition.
In particular, we find that the transition towards
the flocking turbulence can be realized also keeping fixed $\alpha$
and increasing the strength of the self advection
coefficient $\lambda$. 
These observations can be rationalized by considering
the time scales of the different terms of the TTSH model.
The characteristic time scale of the Landau force, 
of the SH term, and of the self-advection term are
$\tau_{\alpha} = 1/|\alpha|$, $\tau_{\Gamma} = \Gamma_4/\Gamma_2^2$ and 
$\tau_{\lambda} = 1/|\alpha|\lambda$, respectively. 
The expression of $\tau_\lambda$ follows from the consideration that, 
while the typical intensity of the velocity $U$ is determined by the 
Landau force, the velocity field in (\ref{eq:1}) is advected by the 
rescaled velocity $\lambda U$. 
In the case of pusher-like swimmers ($\lambda > 1$) 
the self-advection time $\tau_\lambda$ is shorter that $\tau_\alpha$, 
therefore, the transition is expected to occur when
$\tau_\Gamma \sim \tau_\lambda$, i.e., 
\begin{equation}
 |\alpha| \lambda \sim \frac{\Gamma_2^2}{\Gamma_4} \;.
 \label{eq:4}
\end{equation} 
Our prediction generalizes the criterion
proposed in~\cite{wensink2012meso} for the case with constant $\lambda$.
For $\lambda=3.5$ the relation~(\ref{eq:4})
give the transition value $\alpha \simeq -1.14$,
while for fixed $\alpha=-1$ we get $\lambda \simeq 4$,
which are both in quantitative agreement with our numerical findings.
The dependence of the transition on $\lambda$ is an evidence
of the out-of-equilibrium nature of this phenomenon.
This is reminiscent of the properties of the original Toner-Tu model,
where the nonequilibrium nonlinearities are crucial in the establishment
of the broken symmetry state
in two-dimensions~\cite{toner1998flocks,toner2005hydrodynamics,toner2012reanalysis}. 

Finally, we investigated the effects of the confinement in the circular domain.
Our results show that these effects become relevant
when the radius of the domain
is of the order of the correlation scale of the flow.
In the regime of mesoscale turbulence,
the correlation scale $\xi \sim \Lambda$ is much smaller
that the radius $R$ of the domain considered in our study,  
therefore the flow is weakly affected by the confinement.
Strong effects in this regime are expected only in the
case of very small domains with $R \sim \Lambda$.
Conversely, the correlation scale in the regime of flocking
turbulence is large ($\xi \gg \Lambda$),
therefore the flow is influenced by the presence
of the circular boundary. 
We find that the friction with the boundary dissipates
part of the energy
and causes the production of vorticity close to the boundary.
Increasing the confinement, we observe a
spectral condensation of the energy in the lowest accessible mode,
which causes a steepening of the energy spectrum.

Moreover, the confinement can induce a further transition
toward an ordered state of circular flocking with uniform velocity,  
in which the swimmers are organized in a single giant vortex
which spans the whole domain. 
The formation of this state has been previously reported
in~\cite{puggioni2022giant}.
Here we show that this process has a stochastic nature
which manifests in the extreme variability of the transition times.
This is confirmed by the broad power-law tail of the 
probability distribution function of the transition times. 
Investigating the dependence of the statistics of the transition times
on the parameters of the model requires a tremendous computational effort,
which is demanded to future studies. 


\section*{Acknowledgments}
We acknowledge support from the Departments of Excellence grant (MIUR)
and INFN22-FieldTurb.


\bibliography{biblio}

\end{document}